\documentclass[english,twocolumn,superscriptaddress]{revtex4}
\usepackage[T1]{fontenc}
\usepackage[utf8]{inputenc}
\usepackage{graphicx}
\usepackage{amssymb}
\usepackage{amsmath}
\usepackage{placeins}
\usepackage{float}
\usepackage{siunitx} 

\def \bA{{\bf A}}
\def \bk{{\bf k}}

\newcommand{\del}{\partial}
	\newcommand{\D}{\mathrm{d}}
	\newcommand{\dd}[2]{\frac{\mathrm{d} #1}{\mathrm{d} #2}}
	\newcommand{\ddel}[2]{\frac{\del #1}{\del #2}}
	
	\newcommand{\bbm}{\begin{pmatrix}}
	\newcommand{\ebm}{\end{pmatrix}}
	\newcommand{\bma}{\begin{matrix}}
	\newcommand{\ema}{\end{matrix}}
	\newcommand{\bsm}{\begin{smallmatrix}}
	\newcommand{\esm}{\end{smallmatrix}}
	\newcommand{\bsbm}{\left( \begin{smallmatrix}}
	\newcommand{\esbm}{\end{smallmatrix}\right)}
	\newcommand{\To}{\rightarrow}

	\newcommand{\vb}[1]{\left( #1 \right)}
	\newcommand{\vsb}[1]{\left[ #1 \right]}
	
	\newcommand{\ave}[1]{\left\langle #1 \right\rangle}

	\newcommand{\mbf}[1]{\mathbf{#1}}
	\newcommand{\abs}[1]{\vert #1 \vert}

	\newcommand{\dt}{\,\text{.}}

	\newcommand{\ea}{\epsilon_{\mbf k + e \mbf A}}
	\newcommand{\eb}{\delta_{\mbf k + e \mbf A}}

	\newcommand{\usi}[2]{\ensuremath{#1 \, \si{#2}}}

	\newcommand{\com}{\quad\text{,}}
		
	\newcommand{\eqb}{\nonumber\\ &\quad}

	\usepackage[normalem]{ulem} 
	\usepackage{xcolor}
	\definecolor{mgreen}{RGB}{21,105,26}

	\newcommand{\braket}[1]{\ave{#1}}



\usepackage{babel}

\begin{document}

\title{Floquet dynamics in light-driven solids}


\author{M.~Nuske}
\affiliation{Zentrum f\"ur Optische Quantentechnologien, Universit\"at Hamburg, 22761 Hamburg, Germany}
\affiliation{Institut f\"ur Laserphysik, Universit\"at Hamburg, 22761 Hamburg, Germany}
\affiliation{The Hamburg Center for Ultrafast Imaging, Luruper Chaussee 149, Hamburg 22761, Germany}
\author{L.~Broers}
\affiliation{Zentrum f\"ur Optische Quantentechnologien, Universit\"at Hamburg, 22761 Hamburg, Germany}
\affiliation{Institut f\"ur Laserphysik, Universit\"at Hamburg, 22761 Hamburg, Germany}
\author{B.~Schulte}
\affiliation{Max Planck Institute for the Structure and Dynamics of Matter, Luruper Chaussee 149, 22761 Hamburg, Germany}
\author{G.~Jotzu}
\affiliation{Max Planck Institute for the Structure and Dynamics of Matter, Luruper Chaussee 149, 22761 Hamburg, Germany}
\author{S.~A.~Sato}
\affiliation{Center for Computational Sciences, University of Tsukuba, Tsukuba 305-8577, Japan}
\affiliation{Max Planck Institute for the Structure and Dynamics of Matter, Luruper Chaussee 149, 22761 Hamburg, Germany}
\author{A.~Cavalleri}
\affiliation{Max Planck Institute for the Structure and Dynamics of Matter, Luruper Chaussee 149, 22761 Hamburg, Germany}
\author{A.~Rubio}
\affiliation{Max Planck Institute for the Structure and Dynamics of Matter, Luruper Chaussee 149, 22761 Hamburg, Germany}
\affiliation{Center for Computational Quantum Physics (CCQ), Flatiron Institute, 162 Fifth Avenue, New York, NY 10010, USA}
\author{J.~W.~McIver}
\affiliation{Max Planck Institute for the Structure and Dynamics of Matter, Luruper Chaussee 149, 22761 Hamburg, Germany}
\author{L.~Mathey}
\affiliation{Zentrum f\"ur Optische Quantentechnologien, Universit\"at Hamburg, 22761 Hamburg, Germany}
\affiliation{Institut f\"ur Laserphysik, Universit\"at Hamburg, 22761 Hamburg, Germany}
\affiliation{The Hamburg Center for Ultrafast Imaging, Luruper Chaussee 149, Hamburg 22761, Germany}




\begin{abstract}
We demonstrate how the properties of light-induced electronic Floquet states in solids impact natural physical observables, such as transport properties, by capturing the environmental influence on the electrons. We include the environment as dissipative processes, such as inter-band decay and dephasing, often ignored in Floquet predictions. These dissipative processes determine the Floquet band occupations of the emergent steady state, by balancing out the optical driving force. In order to benchmark and illustrate our framework for Floquet physics in a realistic solid, we consider the light-induced Hall conductivity in graphene recently reported by J.~W.~McIver, {\it et al.}, Nature Physics (2020). 
We show that the Hall conductivity is estimated by the Berry flux of the occupied states of the light-induced Floquet bands, in addition to the kinetic contribution given by the average band velocity. Hence, Floquet theory provides an interpretation of this Hall conductivity as a geometric-dissipative effect.
We demonstrate this mechanism within a master equation formalism, and obtain good quantitative agreement with the experimentally measured Hall conductivity, underscoring the validity of this approach which establishes a broadly applicable framework for the understanding of ultrafast non-equilibrium dynamics in solids.
\end{abstract}

\maketitle

	Light control of matter has emerged as a new chapter of condensed matter physics. While the established approach to solid state physics is to probe equilibrium or near-equilibrium properties of a given material, we now take a  more active stance, to design  non-equilibrium states with desired properties by periodic driving. This new vantage point is reflected in recent experimental work on light-controlled superconductivity,  see e.g. \cite{fausti_light-induced_2011,hu_optically_2014, mitrano_possible_2016,okamoto_theory_2016,okamoto_transiently_2017}, where a superconducting state is either enhanced or induced by applying terahertz pulses.
	More generally, optical control provides a dynamical avenue towards creating functionalities on demand in materials \cite{basov_towards_2017}, for which we provide an efficient theoretical framework and understanding.

	A natural theoretical description of a periodically driven system utilizes Floquet theory to determine its quasi-energy states. This approach formally represents a periodically driven, time-dependent Hamiltonian as a time-independent one, which allows the use of time-independent methodologies. If the quasi-energy states are interpreted as the eigenstates of an effective Hamiltonian, this effective Hamiltonian can be qualitatively distinct from the unperturbed Hamiltonian. This approach constitutes 'Floquet engineering' via periodic driving.
	Implementing this approach in nearly-isolated cold atom systems has resulted in spectacular properties \cite{aidelsburger_experimental_2011,tarruell_creating_2012,uehlinger_double_2013,struck_engineering_2013,jotzu_experimental_2014,aidelsburger_measuring_2015,lohse_thouless_2016,grossert_experimental_2016,nakajima_topological_2016}.

	While this approach has a suggestive character, we demonstrate that a naive treatment of the Floquet states as energy states, which are then occupied by electrons with an equilibrium distribution, is in general not a correct prediction for the driven system. 
	Firstly, the measurable properties, such as transport properties, of the driven systems are generally different from the measurable properties of the effective Hamiltonian. The linear response to a probing term in the Hamiltonian, which models the physical probe,  interferes with the driving term. The resulting linear response cannot   be expressed as the linear response of the effective Hamiltonian, in general. 
	Secondly, for a well behaved effective Hamiltonian to describe the low-frequency dynamics, the high-frequency limit of the driving frequency is desirable. Typically that implies that the driving frequency is large compared to the electronic band width, to avoid resonant driving of interband transitions. However, in this high-frequency limit a high driving intensity is required, so that the far off-resonant optical pumping has a noticeable effect on the system. As we point out below, this implies  currently unrealistic experimental and material requirements.
	Thirdly, the steady state of the electrons that emerges in the driven system is in general not an equilibrium distribution on the Floquet quasi-energies. These properties of driven systems emphasize clear distinctions between a system with a Floquet engineered Hamiltonian and a system with a static Hamiltonian. 	

	\begin{figure*}[htb]
		\centering
		\includegraphics[width=\linewidth]{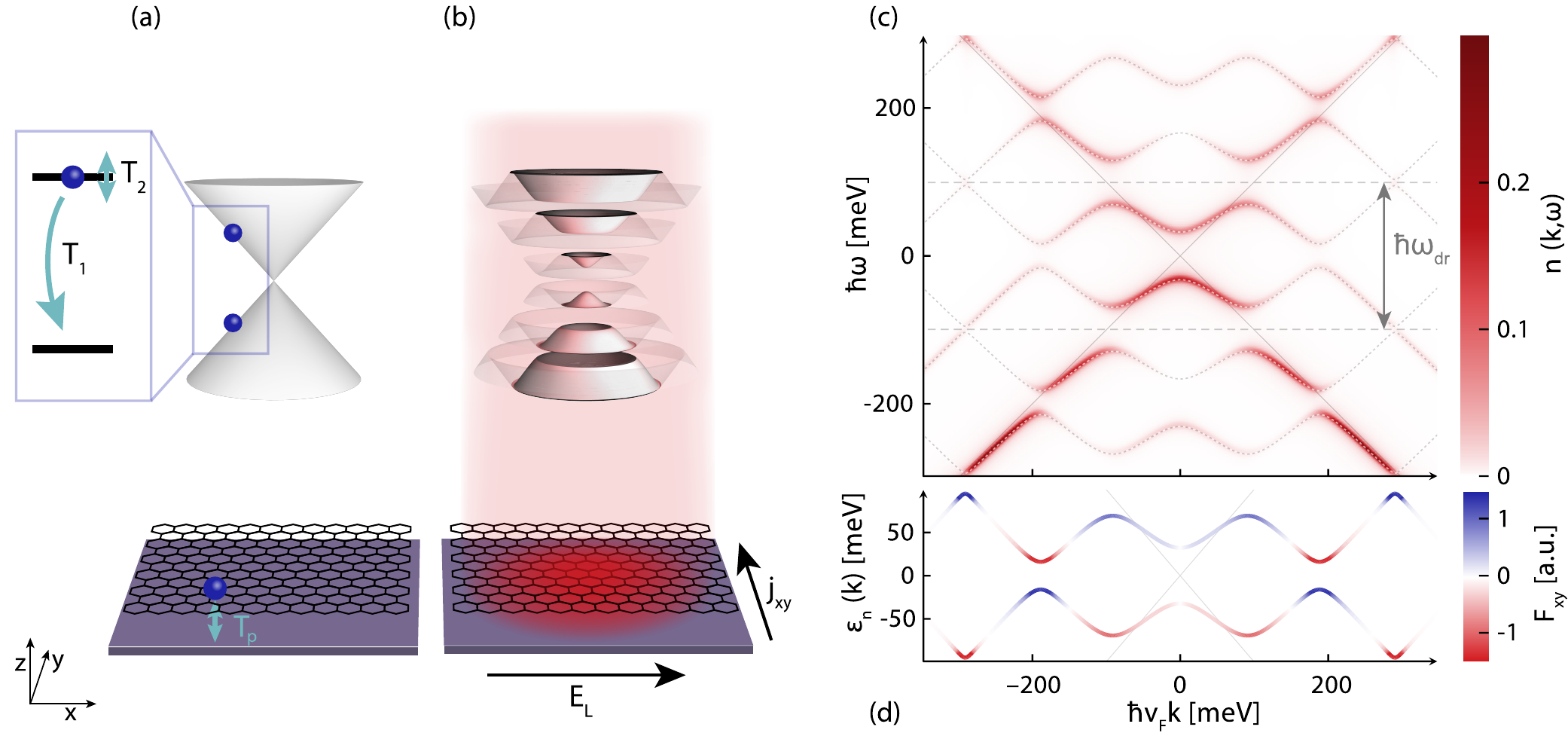}
		\caption{Dirac cone of the undriven (a) and driven (b) graphene band structure (top) and the corresponding real-space lattice (bottom). The band structure of graphene driven with circularly polarized light develops gaps at each resonance and at the Dirac point. Applying a longitudinal field $E_{\rm L}$ induces a transverse Hall current $j_{\rm xy}$.
		(c) Electron distribution $n(\mbf k, \omega)$ as a function of momentum times $\hbar v_{\rm F}$. Note that $\hbar v_Fk=\usi{200}{\milli\eV}$ corresponds to $k\approx\usi{0.03}{\per\angstrom}$. The distribution is shown after a steady state is achieved for a tanh-type ramp to the driven state. The parameters for (c) are inspired by experimental ones used in \cite{mciver_light-induced_2019}.
		Dotted gray lines show the numerically computed Floquet band structure, see App.~\ref{app:floquetBerry}. The maxima of the electron distribution of the driven state agree perfectly with the Floquet band structure. We show a slice along the $k_x$-direction of the band structure shown in panel (b). Dashed gray lines separate the different Floquet replicas. 
		(d) Floquet band structure colored by Berry curvature, for details see App.~\ref{app:floquetBerry}. The Berry curvature is integrated over ring segments of momentum space. This integrated quantity suppresses the curvature at the Dirac point. We only show the first Floquet replica since Berry curvature and Floquet energies are periodic with $\omega_{\rm dr}$. We use $E_{\rm dr}=\usi{26}{\mega\volt\per\meter}$, $\omega_{\rm dr}=2\pi\cdot \usi{48}{\tera\hertz}\approx\usi{200}{\milli\eV}/\hbar$, $T_1=\usi{1}{\pico\second}$, $T_2=\usi{0.2}{\pico\second}$, $T_p=\usi{0.4}{\pico\second}$, temperature $T=\usi{80}{\kelvin}$ and $\mu=0$. Faint gray lines indicate the Dirac cone for undriven graphene.}
		\label{fig:floquetBands}	
	\end{figure*}

	In this paper, we present how the Floquet band properties manifest themselves in transport properties in an optically driven solid. As a central example we consider light control of graphene. 
	References \cite{calvo_tuning_2011,usaj_irradiated_2014,perez-piskunow_floquet_2014,dehghani_dissipative_2014,kristinsson_control_2016} have proposed to illuminate graphene with circularly polarized light with the purpose of inducing a topologically insulating state \cite{oka_photovoltaic_2009,kitagawa_topological_2010,kitagawa_transport_2011,dehghani_out--equilibrium_2015,mikami_brillouin-wigner_2016}, with the same low-energy behavior as the Haldane model \cite{haldane_model_1988}.
	We note that these proposed experiments would only reproduce the behavior of the Haldane model in a band insulating state, under the above mentioned assumption of a large driving frequency. As we demonstrate below, neither of these assumptions is fulfilled. 

	Our primary experimental motivation derives from the measurements of Ref.~\cite{mciver_light-induced_2019}. The authors report on a newly developed on-chip femtosecond technology to detect the Hall current of graphene illuminated with light with a frequency of tens of terahertz, which is orders of magnitude below the band width of graphene. These measurements illustrate the realistic regime of current experiments, and are of guidance for our study. However, we emphasize that our conceptual approach directly applies to any light-driven Dirac material \cite{wehling_dirac_2014,vafek_dirac_2014,zawadzki_semirelativity_2017,armitage_weyl_2018,balatsky_functional_2018}, and more broadly to any solid with well-defined electron-like quasi-particles. A theoretical study on dissipative dynamics in graphene has been reported in Ref.~\cite{sato_microscopic_2019,sato_light-induced_2019}. 

\section*{Geometric-dissipative origin of Hall conductivity}	
	We develop a master equation description for the transport properties of illuminated graphene under realistic conditions. As a key addition to the unitary evolution we include several dissipative processes to provide an effective model for the relaxation and dephasing of the electronic states. These are shown schematically in Fig.~\ref{fig:floquetBands}(a). The form and the magnitude of the dissipative processes determine the steady state that is induced by the optical driving. For this steady state, we determine the Hall current $j_y$ by applying a DC probing field $E_{\rm L}$ along the x-axis. From the linear response definition $\mbf j_y = \sigma_{\rm xy} E_{\rm L}$, we determine the Hall conductivity $\sigma_{\rm xy}$.

	Furthermore, we determine the distribution of electrons in momentum and frequency space $n(\mbf k,\omega)$, which is the Fourier transform of the single-particle correlation function. $n(\mbf k,\omega)$ is depicted in Fig.~\ref{fig:floquetBands}(c). 
	This distribution describes what frequencies and momenta are contained in the time evolution of the electrons, shown here for the steady state. This quantity is closely related to the quantity measured in time- and angle-resolved photoemission spectroscopy (trARPES) experiments \cite{freericks_theoretical_2009}.
	We note that this distribution is consistent with the Floquet bands, depicted in Fig.~\ref{fig:floquetBands}(b) and shown as dashed lines in Fig.~\ref{fig:floquetBands}(c). 
	Furthermore, the Floquet bands are populated primarily in regions that are close to the original Dirac dispersion. This implies that the predominantly occupied Floquet band switches as each resonance. We refer to this property of the band occupation as a split-band picture. This is in contrast to the majority of previous works on Floquet theory where the properties of the continuously connected bands are studied.
	With these observations we determine the derived quantity $n_\sigma(\bk)$,  with band index $\sigma = \pm 1$, which is determined by integrating the distribution $n(\bk,\omega)$ for fixed k in the vicinity of every second frequency maximum, see also App.~\ref{app:singleParticleCorrelation}. 
	This provides an estimate of the occupation of the Floquet states that includes the dissipative broadening of the bands.

	As an additional property of the driven state, we determine the Floquet bands, see Fig.~\ref{fig:floquetBands}(b) and (c). From these, we determine the y-component of the band velocity $v_y^\sigma$ and the Berry Curvature $\Omega^\sigma$. We combine these quantities and define
	\begin{align*}
		\Phi_{\rm xy}&=\frac{1}{A} \sum_{\substack{\mbf k \in {\rm 1. BZ}\\ \sigma=\pm 1}} \Omega^\sigma(\mbf k)  n_{\sigma}(\mbf k) \\
		\bar v_y &= \frac{1}{nA} \sum_{\substack{\mbf k \in {\rm 1. BZ}\\ \sigma=\pm 1}}  v^{\sigma}_{y}(\mbf k)  n_{\sigma}(\mbf k) \com
	\end{align*}	
	where $A$ is the lattice size and $n=1/A \sum_{\bk,\sigma} n_\sigma(\bk)$ is the electron density.
	The Berry flux $\Phi_{\rm xy}$ is the sum over the Berry curvature of the Floquet bands, weighted with the band occupation $n_\sigma(\bk)$. Similarly, the average band velocity $\bar{v}_y$ is the sum over the y-component of the band velocity of the Floquet bands, weighted with the band occupation, and normalized with the electron density.
	The average band velocity term is non-zero for the light-driven state in the presence of the DC field $E_{\rm L}$, due an occupation imbalance along the y-direction, contributing to the Hall current.
	The central result of our study is that the combination of these quantities provides a good estimate of the Hall conductivity
	\begin{align}
		\sigma_{\rm xy}\approx n \bar v_y/E_{\rm L} + \Phi_{\rm xy} \label{eq:relationCurvatureCond}
	\end{align}
	Both the Berry flux and the average band velocity is a sum weighted with the steady state distribution of the driven state, which in turn is determined by the dissipative processes. Our result demonstrates that the Hall conductivity is a geometric-dissipative phenomenon. 
	As we demonstrate below, for small driving field $E_{\rm dr}$, the average band velocity dominates in this prediction, whereas for large driving field, the Berry flux dominates.
	We note that the non-vanishing expectation value of $\bar{v}_y$ derives from an occupation imbalance in the transverse direction of the probe, which was also discussed in Ref.~\cite{sato_microscopic_2019}.

	We note that the Floquet states not only display a topological band gap at the Dirac points, as they would for large driving frequencies, see Refs.~\cite{oka_photovoltaic_2009,kitagawa_topological_2010,kitagawa_transport_2011,dehghani_out--equilibrium_2015,mikami_brillouin-wigner_2016}. In addition to this renormalization of the Dirac cone, additional resonances appear at integer multiples of the driving frequency. Frequency space naturally separates into Floquet zones of the size of the driving frequency, in analogy to Brillouin zones. Each Floquet zone contains two bands, corresponding to the underlying two-band structure of graphene. 
 	Each resulting Floquet band has additional Berry curvature at the resonances, in addition to the curvature at the Dirac point. Integrating over the entire band gives the Chern number of each band. For the example shown, there are about 80 resonances stemming from multi-photon absorption, and the Chern number of the Floquet bands is of the order of $10^2$--$10^3$. However, this is not the magnitude of the Hall conductivity, because the band is not occupied in a band insulating state, but rather has the electron distribution depicted in Fig.~\ref{fig:floquetBands}(c). 
 	For this example, we find that $99.8\%$ of the Hall conductivity can be accounted for by summing the contributions from the Dirac point and the first four resonances. 
 	We note that the total Hall current has the opposite sign of that expected in the high-frequency limit. We observe that the Hall conductivity is not quantized in an obvious fashion, however, we find a soft plateau of the conductivity as a function of the driving field, more pronounced when depicted as a function of fluence, see App.~\ref{app:fluenceDependence}. While the magnitude of the conductivity at the plateau depends on the model assumptions, such as the choice of dissipative processes, the robustness of this feature might point to an underlying principle to be discussed elsewhere. 

\section*{Rabi solution}
 	The key qualitative difference to the proposals that utilize high-frequency driving, is the occurrence of resonances at integer multiples of the driving frequency. As depicted in Fig.~\ref{fig:floquetBands}(d), these resonances create Berry curvature of the Floquet bands. 
	To demonstrate this point we determine the Berry curvature at the single-photon resonance within a Rabi picture, which gives access to all properties near the single-photon resonance and provides analytical expressions. We expand on this analysis in a Floquet picture further down, which treats the full system within a numerical framework. We describe the graphene dispersion and the interaction of the electrons with the electromagnetic field via 
	\begin{align*}
		H_{R}&=\hbar v_{\rm F} k \sigma_z + \frac{\sigma_{\rm pol} e E_{\rm dr}v_{\rm F}}{2 \omega_{\rm dr}} \bbm 0 \qquad ie^{-i \omega_{\rm dr}t-i \tau_z \sigma_{\rm pol} \phi_k}  \\  -ie^{i \omega_{\rm dr}t+i \tau_z \sigma_{\rm pol} \phi_k} \qquad 0 \ebm 	
	\end{align*}
	and hence the solutions $|\psi_{R,\pm}(t)\rangle$ can be obtained in analogy to the Rabi problem, see App.~\ref{app:rabi}. Here $k e^{i\phi_k}=k_x+ik_y$ and $\tau_z=\pm1$ labels the two inequivalent Dirac points, $\sigma_{\rm pol}=\pm1$ determines the polarization of the light, $v_{\rm F}$ denotes the Fermi velocity and $e>0$ is the elementary charge. We then compute the instantaneous Berry curvature within the Rabi approximation, which gives
	\begin{align}
		\Omega^\pm(\mbf k)&=\mp \frac{e^2}{h}\,\frac{\sigma_{\rm pol}v_{\rm F} \lambda^2}{2k\Omega_R^3} \label{eq:RabiBerryCurv} \com
	\end{align}
	where $\lambda= \frac{e E_{\rm dr} v_{\rm F}}{2 \omega_{\rm dr}}$ is the bare Rabi frequency, $\Delta=\frac{\omega_{dr}-2v_{\rm F} k}{2}$ is the detuning and $\Omega_R=\sqrt{\lambda^2+\Delta^2}$ is the Rabi frequency.
	We emphasize that this result applies directly to any light-driven Dirac material, and that similar considerations can be extended naturally to any material with electronic quasi-particles. 
	As we discuss below, a Floquet analysis expands this analysis to all resonances numerically.
			
	\begin{figure}[tb]
		\centering
		\includegraphics[width=\linewidth]{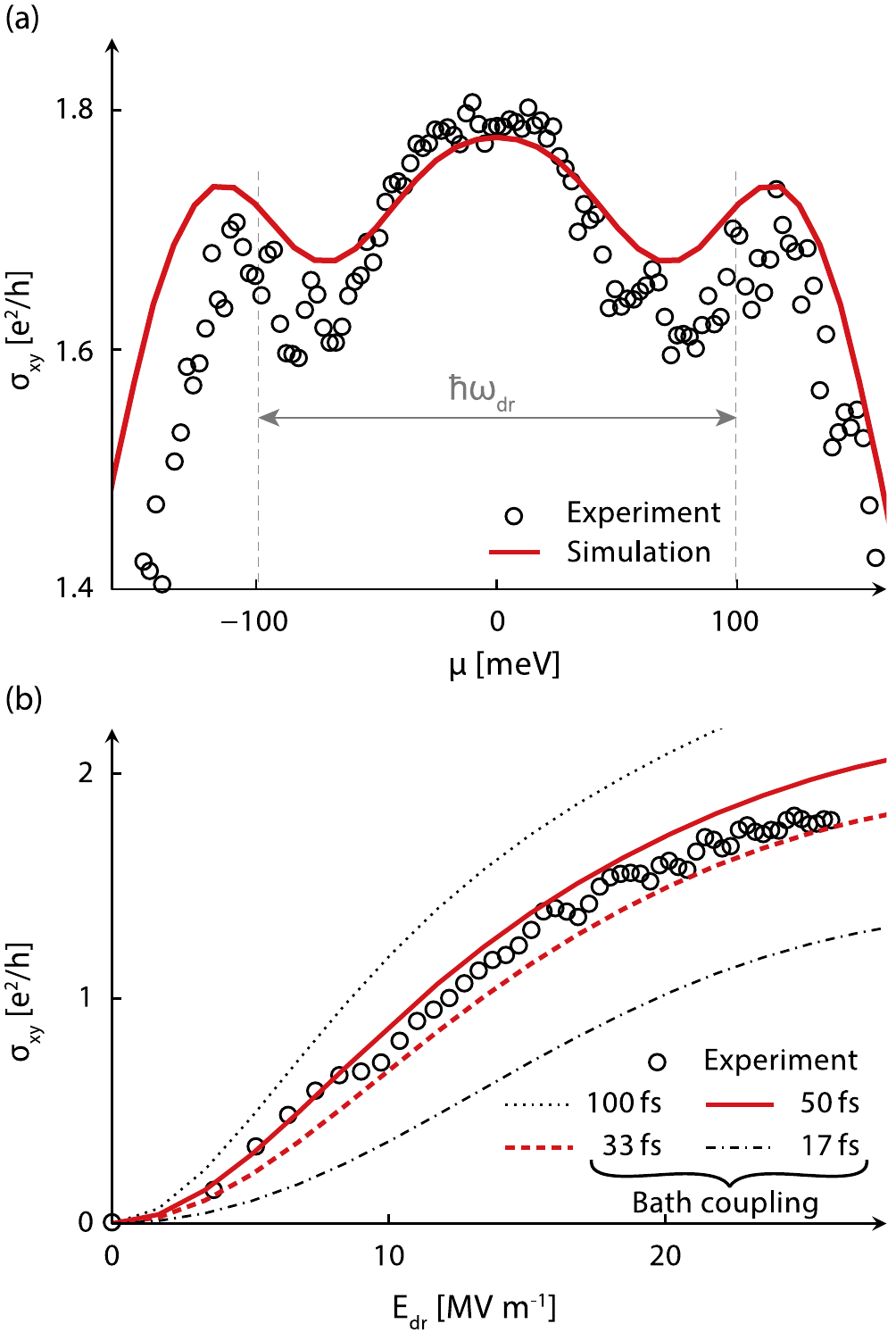}
		\caption{(a) Circular dichroism of the transverse conductivity as a function of chemical potential. The data from the numerical simulation (red line) and experimental data \cite{mciver_light-induced_2019} (open circles) agree quantitatively. (b) Electric-field dependence of the current dichroism for several values of the particle-exchange time-scale $T_p$, as indicated in the legend. We see that a value in the range of $T_p=\usi{30-50}{\femto\second}$ is consistent with the experiment. The parameters for the numerical simulation are $\omega_{\rm dr}=\usi{2\pi\cdot 48}{\tera\hertz}\approx\usi{200}{\milli\eV}/\hbar$, $T_1=\usi{100}{\femto\second}$, $T_2=\usi{20}{\femto\second}$, $T=\usi{80}{\kelvin}$, $E_{\rm L}=\usi{1.7}{\kilo\volt\per\meter}$ and the driving pulse has a Gaussian envelope with electric-field strength FWHM of $\usi{\sqrt 2}{\pico\second}$, corresponding to intensity FWHM of $\usi{1}{\pico\second}$. Finally $E_{\rm dr}=\usi{26}{\mega\volt\per\meter}$ and $T_p=\usi{36}{\femto\second}$ in panel (a) and chemical potential $\mu=0$ in panel (b).}
		\label{fig:highFluenceCompExp}	
	\end{figure}

\section*{Master equation}

	To evaluate the quantity $n \bar v/E_{\rm L} + \Phi_{\rm xy}$ of Eq.~\ref{eq:relationCurvatureCond}, we determine the steady state occupation $n_\sigma(\bk)$ numerically. Similarly, we determine the Hall conductivity.
	We factorize the density matrix $\rho$ of the system as $\rho = \prod_\bk \rho_\bk$, where we choose a discrete lattice of momenta $\bk$, centered around the Dirac point, of size $N\times N$.
	We represent each $\rho_\bk$ in the four-dimensional basis of $|0\rangle$, $c_{\bk,+}^\dagger |0\rangle$, $c_{\bk,-}^\dagger |0\rangle$, $c_{\bk,+}^\dagger c_{\bk,-}^\dagger |0\rangle$. The operators $c_{\bk,\sigma}$ describe the upper and the lower band of momentum $\bk$, depicted in Fig.~\ref{fig:floquetBands}(a). We note that using this four-dimensional basis enables us to determine the electron distribution $n(\bk,\omega)$, and treat undoped and doped graphene in a systematic manner, by varying the chemical potential.
	For each $\rho_\bk$, we solve the master equation.
	The master equation contains unitary contributions from the equilibrium Hamiltonian $H_0$ and the light-matter interaction $H_{\rm em}$. The latter contains both the circularly polarized driving term, with electric-field strength $E_{\rm dr}$, as well as a longitudinal DC probing field $E_{\rm L}$.  
	In addition to these unitary contributions, we introduce dissipative processes, depicted in Fig.~\ref{fig:floquetBands}(a), modeled via Lindblad operators, see App.~\ref{app:numericalDetails}. The first describes decay from the upper to the lower band, with rate $\gamma_{1}=1/T_{1}$. The second describes dephasing between the upper and the lower band, described by a rate $\gamma_{z}$, which we combine into $\gamma_2= 1/T_{2}=1/(2T_1)+2\gamma_z$. 
	The third rate $\gamma_{p} = 1/T_{p}$ corresponds to single-particle exchange with a fermionic bath of temperature $T$ and chemical potential $\mu$.

	We first demonstrate that the experimental results of Ref.~\cite{mciver_light-induced_2019} are captured with this model.
	In Fig.~\ref{fig:highFluenceCompExp} we compare the circular dichroism of the Hall conductivity, which is defined as one half the difference of the response for right- and left-handed circular polarization, of the measurement and our calculation. 
	We find that both the peak-field dependence in Fig.~\ref{fig:highFluenceCompExp}(b) as well as the chemical potential dependence for high fluence in Fig.~\ref{fig:highFluenceCompExp}(a) are in quantitative agreement.
	The chosen $E_{\rm dr}$ and $\omega_{\rm dr}$ correspond to the peak driving field and central frequency of the laser pulses used in the experiment, respectively \cite{mciver_light-induced_2019}. The dissipation rate $T_1$ is inspired by Ref.~\cite{gierz_snapshots_2013} and for $T_2$ we choose \usi{20}{\femto\second}, which is chosen to be notably smaller than $T_1$. The decay rate $T_{p}$ is adjusted to match the experimental data. We find that $T_{\rm p}=\usi{30-50}{\femto\second}$ are appropriate depending on the electric field strength, see Fig.~\ref{fig:highFluenceCompExp}(b).
	We emphasize that the properties of the driven state crucially depend on the dissipative environment. Both the measurable properties, such as the transport behavior, and the steady state itself are shaped by the dissipation. This key result demonstrates the urgency of including the dissipative environment to model a material, and provides guidance for the design of light-induced material properties. On the conceptual side, it is this dissipative environment that is not captured in a Floquet analysis, but profoundly alters the physical behavior of the system.

	\begin{figure}[b!]
		\centering
		\includegraphics[width=0.95\linewidth]{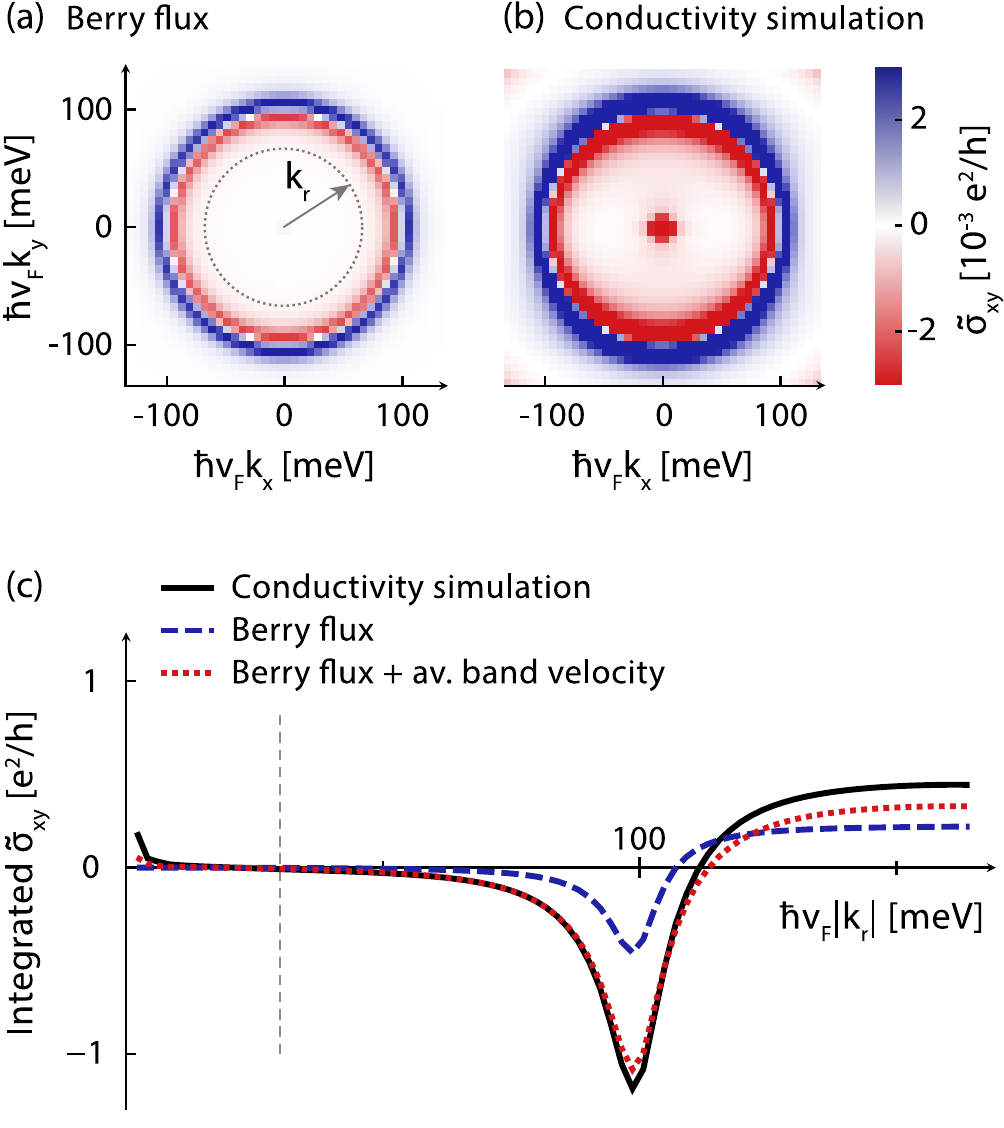}
		\caption{Comparison of the momentum resolved Berry flux, depicted in panel (a), which is obtained within the Rabi approximation, and the momentum resolved conductivity dichroism, depicted in panel (b). The Berry flux displays a qualitatively similar behavior as the conductivity dichroism, in particular a sign change at the single photon resonance. The contribution near the Dirac point is not included in the Rabi approximation.
		(c) We depict the conductivity density, integrated over a disk in momentum space of radius $\bk_{\rm r}$, as shown in panel (a). For comparison, we show the momentum resolved Berry flux, integrated over the same disk in momentum space, and the sum of the integrated Berry flux and the integrated average band velocity.
		The curves have been shifted such that their value vanishes at $\hbar \rm v_F \abs{k_{\rm r}}=\usi{30}{\milli\eV}$ for a better comparison of the contribution of the first resonance. 
		Consistent with the proposed estimate in Eq.~\ref{eq:RabiBerryCurv}, the sum of the Berry flux and the average band velocity predict the conductivity, even in a momentum resolved manner. 
		In all plots we use $E_{\rm dr}=\usi{3}{\mega\volt\per\meter}$, $\omega_{\rm dr}=\usi{2\pi\cdot 48}{\tera\hertz}\approx\usi{200}{\milli\eV}/\hbar$, $T_1=\usi{50}{\pico\second}$, $T_2=\usi{10}{\pico\second}$, $T_p=\usi{20}{\pico\second}$, $T=\usi{80}{\kelvin}$, $E_{\rm L}=\usi{0.84}{\kilo\volt\per\meter}$, $\mu=0$ and the driving pulse is ramped with a tanh over $\usi{1}{\pico\second}$.}
		\label{fig:rabiBerryCurv}
	\end{figure}

\section*{Berry flux of Rabi states}

	In Fig.~\ref{fig:rabiBerryCurv}(b) we depict the contributions to the Hall conductivity in momentum space $\tilde \sigma_{\rm xy}(\bk)=\sigma_{\rm xy}(\bk)/A$, as defined in App.~\ref{app:numericalDetails}. In addition to the negative contributions near the Dirac point, which are not captured by the Rabi approach, there are negative contributions below the single photon resonance, and positive contributions above the resonance. 
	For comparison we depict the contributions to the Berry flux $\Phi_{\rm xy}$ as determined within the Rabi approximation, in Fig.~\ref{fig:rabiBerryCurv}(a). We find that the Rabi solution for the curvature gives a qualitatively correct description of the momentum resolved conductivity. We note that the Rabi solution does not capture two-photon processes, which create the gap opening at the Dirac point, as well as higher order gaps. 

	In Fig.~\ref{fig:rabiBerryCurv}(c) we depict a quantitative comparison. We show the momentum-resolved conductivity contributions integrated over a disk of radius $k_r$, i.e.~we show $\sum_{|k| < k_r} \tilde\sigma_{\rm xy}(\bk)$. Similarly, we show the contributions to the Berry flux $\Phi_{\rm xy}$ integrated to $k_r$, as well as the sum of the curvature and band velocity contributions $\Phi_{\rm xy}+n\bar v/E_{\rm L}$, integrated up to $k_r$. We note that the integrated conductivity has been shifted up such that it has a zero crossing at $\hbar v k_r \approx \usi{30}{\milli\eV}$, so that the behavior at the single photon resonance can be compared directly to the Rabi solution.

	We find that the momentum-resolved representation of $n\bar v_y/E_{\rm L} + \Phi_{\rm xy}$ gives a good prediction for the momentum-resolved conductivity. Generally, the agreement is good for small dissipation, in particular for small $\gamma_z$. The total value of the conductivity, which is the measurable conductivity of the system, is positive, and therefore of opposite sign than the contributions near the Dirac point. This implies that the positive contributions above the resonance, i.e.~momentum states with $v_{\rm F}k>\omega_{\rm dr}/2$, exceed the negative contributions below the resonance, i.e.~$v_{\rm F}k<\omega_{\rm dr}/2$. 
	The sign change of the contributions is a direct consequence of the split-band picture depicted in Fig.~\ref{fig:floquetBands}(c). As mentioned above, in this picture the predominantly occupied band switches at each resonance.
	The specific value of the conductivity depends continuously on the driving frequency and the dissipative properties of the system. 

	The momentum resolved Berry flux, which is depicted in Fig.~\ref{fig:rabiBerryCurv}(a), derives from the Berry curvature and the occupation of the Floquet bands. The Rabi approximation describes the Floquet bands near the single photon resonance. In Fig.~\ref{fig:floquetBands}(c) and \ref{fig:floquetBands}(d), this resonance occurs at $\hbar v_{\rm F} k \approx \pm \usi{100}{\milli\eV}$. The Rabi approximation of the curvature, given in Eq.~\ref{eq:RabiBerryCurv}, predicts negative values of the curvature for the upper band, and positive curvature for the lower band, localized near the resonance. The occupation of the upper band $n_+(\bk)$ is larger than the occupation of the lower band $n_-(\bk)$ for momenta smaller than the resonance. For momenta larger than the resonance we have $n_+(k) < n_-(k)$. This change of predominant occupation results in a partial cancellation of the Berry flux. However, the lower-band contribution dominates, resulting in a positive contribution for the flux. 
	Both the Berry curvature and the Berry flux are rotationally symmetric. In contrast, the average band velocity is manifestly anisotropic, since the band velocity $v_y^\sigma$ vanishes along the $k_x$-direction, see also Ref.~\cite{sato_microscopic_2019}. This gives rise to the modest anisotropy of the Hall conductivity, see Fig.~\ref{fig:rabiBerryCurv}(b). 

	\begin{figure}[tb]
		\centering
		\includegraphics[width=\linewidth]{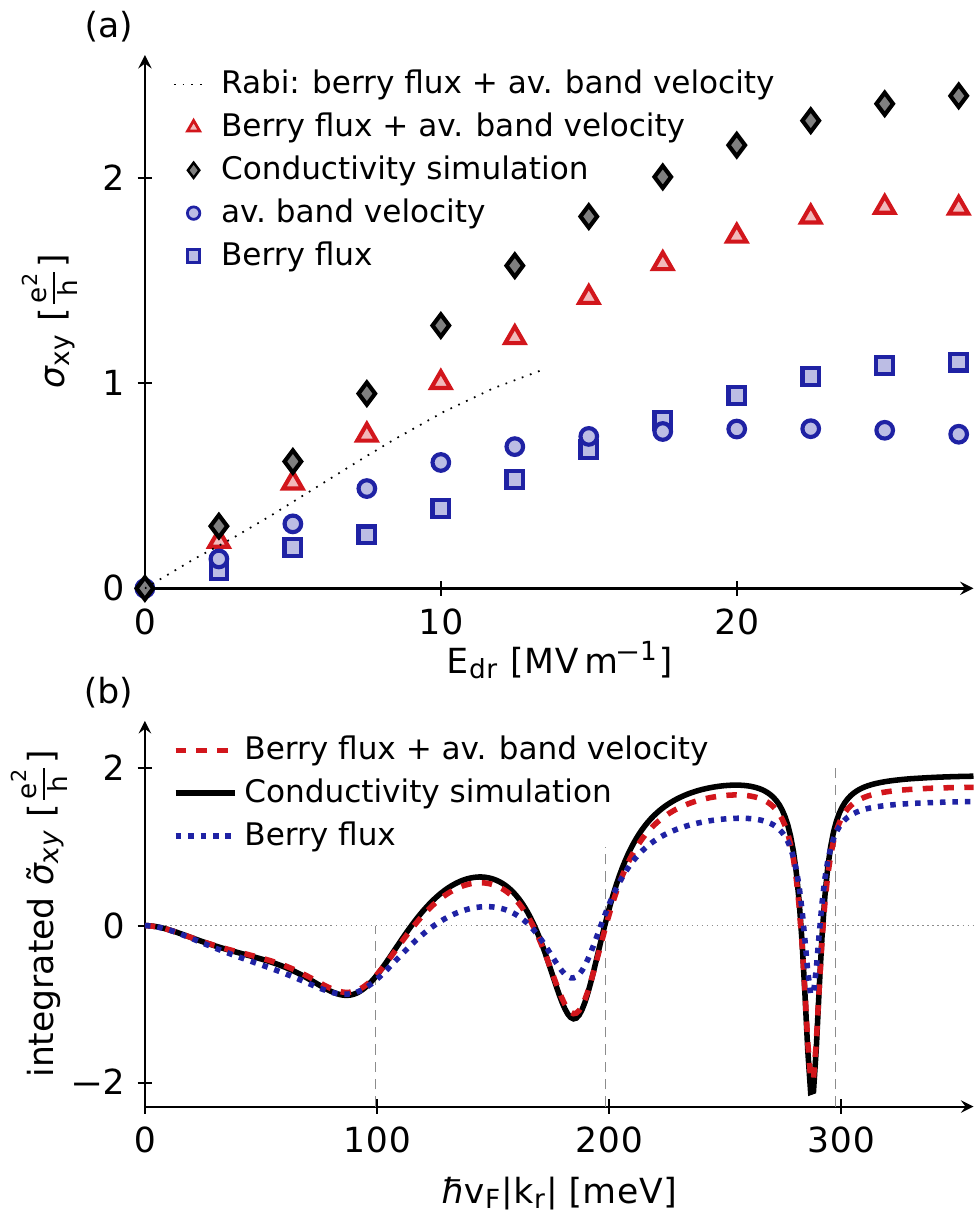}
		\caption{(a) Comparison of the electric-field dependence of the conductivity dichroism, the Berry flux and the averaged band velocity. Black diamonds show the simulated conductivity dichroism, the faint dashed line the sum of the Berry flux and the average band velocity within the Rabi approximation.
		Red triangles show the sum of the Berry flux and the average band velocity, based on Floquet states.
		Blue squares show the Berry flux only and blue circles the average band velocity only. 
		(b) We show the conductivity density $\tilde \sigma_{\rm xy}$, integrated over a disk in momentum space with radius $\bk_r$, for zero dephasing rate $\gamma_z=0$. 
		This integrated conductivity agrees with the sum over the Berry flux and the average band velocity, integrated over the same disk.
		For both panels the parameters are $\omega_{\rm dr}=\usi{2\pi\cdot 48}{\tera\hertz}\approx\usi{200}{\milli\eV}/\hbar$, $T_1=\usi{1}{\pico\second}$, $T_p=\usi{0.4}{\pico\second}$, $T=\usi{80}{\kelvin}$ and $E_{\rm L}=\usi{1.7}{\kilo\volt\per\meter}$. In panel (a) $T_2=\usi{0.2}{\pico\second}$ and in panel (b) $T_2=\usi{2}{\pico\second}$ and $E_{\rm dr}=\usi{26}{\mega\volt\per\meter}$. All observables are shown after a steady state is achieved for a tanh-type ramp of the driving field strength.}
		\label{fig:compFloquetFluenceDep}
	\end{figure}	

\section*{Berry flux of Floquet states}

 	We expand this analysis by determining the Floquet bands of the driven system, and their band velocity and curvature, as described in App.~\ref{app:floquetBerry}. While the Rabi solution gives access to the properties of the single-photon resonance, the Floquet analysis gives the light-induced band properties to any order. We utilize the band velocities and the Berry curvature that is obtained from the Floquet bands, and combine them with the band occupations derived from $n(\bk, \omega)$, as shown in Fig.~\ref{fig:floquetBands}(c), to determine the average band velocity and the Berry flux. 

	In Fig.~\ref{fig:compFloquetFluenceDep}(a), we display these quantities, and the sum of the average band velocity and the Berry flux. We find again that the sum of the Berry flux and the average band velocity gives a good prediction for the conductivity. The prediction is particularly good for small dissipation.
	We also display the Rabi approximation, which gives a good estimate at small $E_{\rm dr}$. We note that for small $E_{\rm dr}$ the band velocity contribution dominates, whereas for larger values of $E_{\rm dr}$, the Berry flux dominates. The Berry flux dominated regime is achieved in the strongly driven regime, because the Floquet bands become flat, and the band velocities throughout the bands approach zero.

	In Fig.~\ref{fig:compFloquetFluenceDep}(b) we display the momentum resolved contributions to the conductivity for zero dephasing rate $\gamma_z=0$. 
	Our prediction for the conductivity based on Berry flux and averaged band velocity agrees almost perfectly with the simulated conductivity. 
	We note that we find equally good agreement for non-zero $\gamma_z$, when considering only the momentum modes along the $k_y$-direction, see App.~\ref{app:conductivityYCut}.
	When considering all momenta and non-zero $\gamma_z$, the contributions to $n\bar v/E_{\rm L} + \Phi_{\rm xy}$ deviate from the contributions to the conductivity $\sigma_{\rm xy}$, giving rise to the deviation between $n\bar v/E_{\rm L} + \Phi_{\rm xy}$ and $\sigma_{\rm xy}$ in Fig.~\ref{fig:compFloquetFluenceDep}(a). This suggests an additional contribution due to the dephasing rate $\gamma_z$, possibly related to coherences between the Floquet bands, to be discussed elsewhere.
	We note that at integer multiples of the resonance frequency at $2 v_{\rm F} k \approx n \omega_{\rm dr}$, the momentum resolved sum of the Berry flux and average band velocity changes sign. This behavior was described for the single photon resonance above, and repeats itself for higher orders. We observe that while the momentum resolved contribution to all three quantities is large, there is a near-cancellation of these contributions for higher order resonances. 

\section*{Conclusion}	
	The conceptual achievement that we put forth here, is widely applicable for the description of light-induced dynamics in solids with well-defined electronic quasi-particles. We have presented a versatile and efficient master equation approach that includes the dissipative environment, enabling the description of light-driven solids under realistic conditions. The dissipative environment, which is ignored in the Floquet description of the driven system, shapes the emerging steady state by balancing out the light-induced force on the electrons. Furthermore, our approach is well suited to describe realistic driving frequencies that are small compared to the electronic bandwidth, and therefore induce resonant inter-band excitations, and treat the dynamics that are induced by probing processes explicitly. 

	Even though the construction of the light-induced Floquet states is an incomplete description of a light-driven solid, because the dissipative environment is ignored, we point out what features of Floquet states manifest themselves in its presence, resulting in Floquet physics in realistic materials. The key elements of our approach were exemplified for the recently observed light-induced Hall effect in graphene, for which we obtain a quantitative understanding. We have shown that the Hall conductivity is predicted by the sum of the average band velocity and the Berry flux of the light induced Floquet bands. Therefore our prediction combines geometric properties of the Floquet bands, and dissipative properties of the material, which identify the Hall effect as a geometric-dissipative effect. This insight, derived from our master equation description, demonstrates the effectiveness of our approach, and motivates its application to a wide range of light-induced dynamics in solids.

\acknowledgments{
We acknowledge support from the Deutsche Forschungsgemeinschaft through the SFB 925. 
This work is supported by the Cluster of Excellence 'CUI: Advanced Imaging of Matter' of the Deutsche Forschungsgemeinschaft (DFG) - EXC 2056 - project ID 390715994.
The research leading to these results received funding from the European Research Council under the European Union’s Seventh Framework Programme (FP7/2007–2013)/ERC Grant Agreement No. 319286 (QMAC).  
M.N.~acknowledges support from Stiftung der Deutschen Wirtschaft. 
AR  acknowledge supported by the European Research Council (ERC-2015-AdG694097) and the Flatiron Institute of the Simons Foundation.
}

\clearpage

\begin{appendix}

	\section{Numerical Algorithm for the computation of currents} \label{app:numericalDetails}
		We use the von Neumann equation for the unitary part of the time evolution and include interactions as well as other damping and dephasing effects by including Lindblad operators. When using the Weyl gauge the Hamiltonian does not couple different momentum points. As mentioned in the main text we therefore consider the ansatz $\rho=\prod_{\bk} \rho_{\bk}$ for the density matrix $\rho$ of the system. The full time evolution of the density matrix is then governed by the master equation \cite{pearle_simple_2012}
		\begin{align*}
			\dd{}{t}\rho_{\mbf k} &= \frac{i}{\hbar} \vsb{\rho_{\mbf k},H_{\mbf k}}\eqb
			-\frac 12 \sum_\alpha \vb{L^{\alpha \dag}L^\alpha \rho_{\mbf k} + \rho_{\mbf k} L^{\alpha\dag}L^\alpha -2 L^\alpha \rho_{\mbf k} L^{\alpha\dag}} \dt
		\end{align*}
		The first line of this equation describes the unitary part of the time evolution, fully determined by the Hamiltonian of the system. The second line with the Lindblad operators $L^\alpha$ accounts for damping and dephasing effects.		

\begin{figure}[b]
			\centering
			\includegraphics[width=0.9\linewidth]{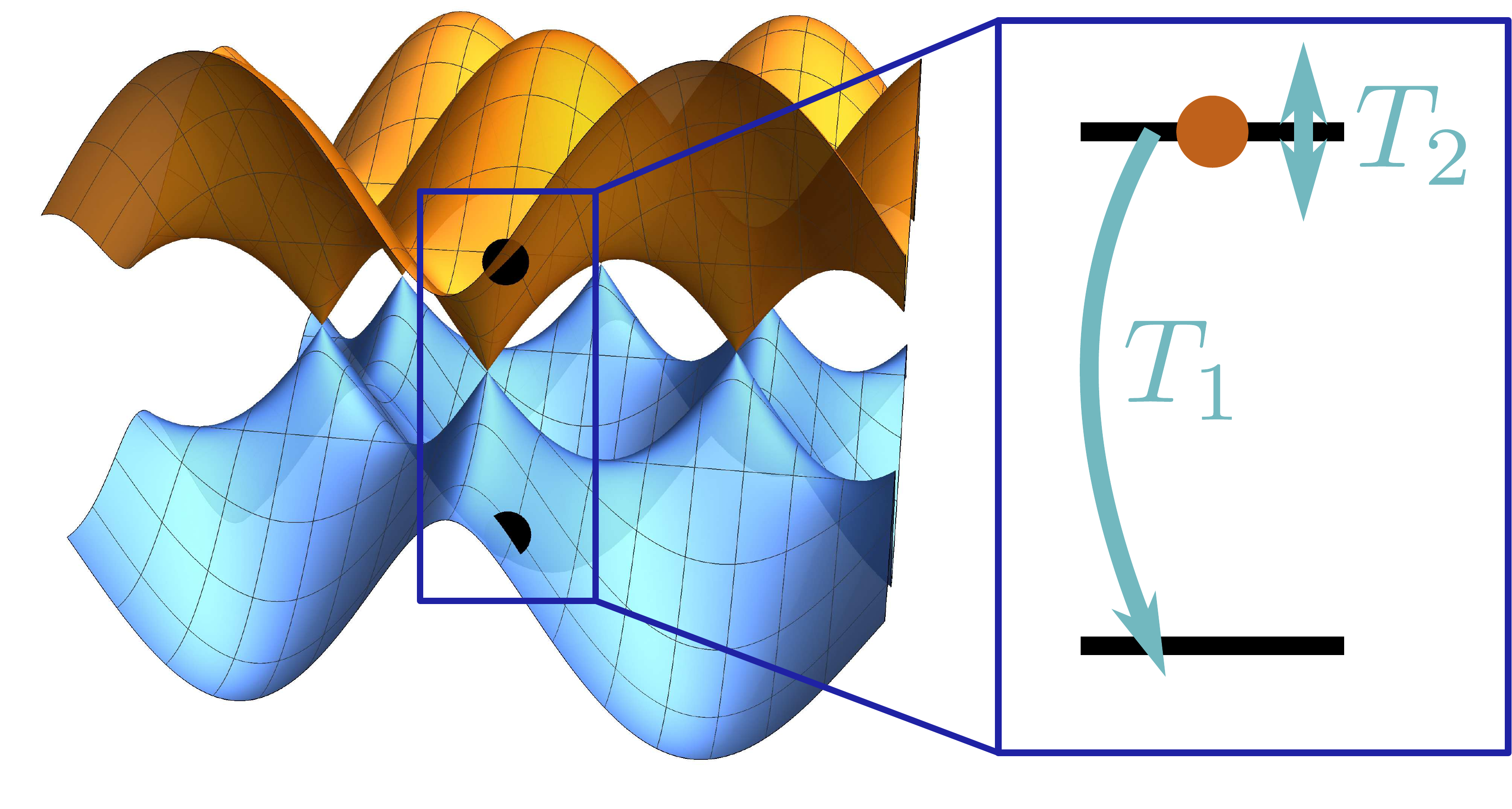}
			\caption{Sketch of the graphene dispersion relation (left). In the Weyl gauge the Hamiltonian decouples in momentum space such that we can treat each momentum point as a two level system (Sketch on right hand side). We also sketch the effect of damping ($T_1$) and dephasing ($T_2$) effects.}
			\label{fig:twoLevelSystem}
		\end{figure}

		In the Weyl gauge (for details see App.~\ref{app:gaugeChoice}) the graphene Hamiltonian with light field coupled via minimal coupling can be written such that it remains diagonal in momentum space
		\begin{align*}
			H&=\sum_{\mbf k} H_{\mbf k}\dt
		\end{align*}		
		Each momentum mode is modeled by a two-level system (see Fig.~\ref{fig:twoLevelSystem}) and hence there are four possible fermionic states, that correspond to an empty mode, a particle on the A sublattice, a particle on the B sublattice and a fully occupied mode
		\begin{align*}
			\Psi_{\mbf k}&= \bbm |11\rangle & |01\rangle & |10 \rangle & |00\rangle \ebm \com
		\end{align*}
		where $|11\rangle=c_{\bk,A}^\dagger c_{\bk,B}^\dagger |0\rangle$ and $c_{\bk,\sigma}^\dag$ creates an electron with momentum $\bk$ in band $\sigma$.

		In order to write the Hamiltonian with respect to this basis we introduce a set of Pauli-type matrices
		\begin{align*}
			\sigma_x&= \bbm 0&0&0&0 \\ 0&0&1&0 \\ 0&1&0&0 \\ 0&0&0&0   \ebm & \sigma_y&= \bbm 0&0&0&0 \\ 0&0&-i&0 \\ 0&i&0&0 \\ 0&0&0&0   \ebm	\\		
			\sigma_z&= \bbm 0&0&0&0 \\ 0&1&0&0 \\ 0&0&-1&0 \\ 0&0&0&0  \ebm & \sigma_z^{(0)}&= \bbm 0&0&0&0 \\ 0&0&0&0 \\ 0&0&1&0 \\ 0&0&0&-1   \ebm \\
			\sigma_z^{(2)}&= \bbm 1&0&0&0 \\ 0&-1&0&0 \\ 0&0&0&0 \\ 0&0&0&0   \ebm & \sigma_g&= \bbm 0&0&0&0 \\ 0&1/2&0&0 \\ 0&0&1/2&0 \\ 0&0&0&0   \ebm
		\end{align*}		

		For the linearized dispersion relation the Hamiltonian for each momentum point further splits into a sum of two terms
		\begin{align}
			H_{\mbf k}&=H_{0,\mbf k}+H_{\rm em}(t)\label{eq:grapheneHamiltonian} \\
			{\rm where}\quad H_{\rm em}(t)&= H_{\rm dr,\mbf k}(t)+H_{L,\mbf k}(t) \dt
		\end{align}
		
		The first contribution is the equilibrium Hamiltonian without any light field applied. Except for the chemical potential it contains only terms of type $\sigma_{x,y,z}$ because the empty and the fully occupied sector do not have a unitary time evolution. In our calculations we include the chemical potential in $H_{0,\bk}$, such that it becomes
		\begin{align*}
			H_{0,\mbf k}&= \Psi_k^\dag \Big[ \hbar v_{\rm F}\vb{ \tau_z k_x \sigma_x + k_y \sigma_y} \eqb -\mu \vb{1+\sigma_z+\sigma_z^{(0)}+\sigma_z^{(2)}} \Big] \Psi_k  \dt
		\end{align*}
		Here $\tau_z=\pm 1$ is the valley index, describing the two Dirac points, $\bk$ is the momentum relative to the Dirac momenta ${\bf K}$ and ${\bf K}'$ and $v_{\rm F}\approx \usi{10^6}{\meter\per\second}$ is the Fermi velocity. We suppress spin indices.

		The second and third terms represent the two light fields, that are coupled through the Peierls substitution $\mbf k \To \mbf k - \frac q \hbar \mbf A(\mbf r,t)$, with charge $q=-{\rm e}$ and ${\rm e}>0$ being the elementary charge. The second term resembles the experimental driving or pump pulse and is a circularly polarized electromagnetic field propagating along the z-direction
		\begin{align*}
			H_{\rm dr,\mbf k}(t)&= ev_{\rm F} \Psi_k^\dag \vb{ \tau_z A_{\rm dr,x} \sigma_x + A_{\rm dr, y} \sigma_y } \Psi_k
		\end{align*}
		where
		\begin{align*}
			{\mbf A}_{\rm dr}&=\bbm A_{\rm dr,x}\\A_{\rm dr,y} \ebm = -\int_0^t \D t' \; {\mbf E}_{\rm dr}(t) \com \\
			\mbf E_{\rm dr}(t)&= - E_{\rm dr}\, g_{\rm env}(t) \vb{\cos(\omega_{\rm dr} t) \mbf e_x + \sigma_{\rm pol} \sin(\omega_{\rm dr} t) \mbf e_y } \com
		\end{align*}
		$\sigma_{\rm pol}$ defines the polarization of the light, $\mbf e_{x,y}$ are unit vectors in $x$- and $y$-direction and $g_{\rm env}(t)$ is the envelope of the pulse. The envelope $g_{\rm env}(t)$ is either chosen to be a Gaussian envelope or a tanh-type switch on. Furthermore we only give the fields in the x-y-plane (at $z=0$) as, without loss of generality, we choose the graphene sheet to lie in this plane. 

		For $g_{\rm env}(t)=1$ we obtain
		\begin{align}
			H_{{\rm dr}, \bk} &=\frac{eE_{\rm dr} v_{\rm F}}{\omega_{\rm dr}} \eqb \Psi_\bk^\dag \vb{ \tau_z \sin(\omega_{\rm dr}t) \sigma_x -\sigma_{\rm pol}  \cos(\omega_{\rm dr}t) \sigma_y} \Psi_\bk \label{eq:pumpHamiltonian} \com
		\end{align}	

		The third term is a direct-current, longitudinal field, that resembles the experimental probe field
		\begin{align*}
			H_{\rm L, \mbf k}&= e v_{\rm F} s_{\rm switch}(t) \Psi_k^\dag \vb{\tau_z  A_{\rm L,x} \sigma_x + A_{\rm L, y} \sigma_y } \Psi_k \com			
		\end{align*}	
		where
		\begin{align*}
			\mbf E_{\rm L}(t)&= E_{L}\mbf e_x \com \\			
			\mbf A_{\rm L}(t)&= -\int_0^t \D t'\; E_{\rm L}(t) = - E_{\rm L} t\, \mbf e_x 
		\end{align*}
		and $s_{\rm switch}(t)$ denotes a switch-on during the first $\usi{0.1}{\pico\second}$. 

		In the high-frequency limit for the pump pulse the second term of the Hamiltonian can be approximated by an effective Hamiltonian describing the low-frequency dynamics of the system\cite{oka_photovoltaic_2009,kitagawa_topological_2010}
		\begin{align*}
			H_{\rm eff,\mbf k}&=-g_{\rm env}(t)\, {\sigma_{\rm pol} \Delta_{\rm hf}} \; \Psi_k^\dag \sigma_z \Psi_k \com
		\end{align*}
		where $\Delta_{\rm hf}=\vb{\hbar v_{\rm F} e E_{\rm dr}}^2/{(\hbar \omega_{\rm dr})^{3}}$.

		In addition to the unitary time evolution governed by the Hamiltonian $H_\bk$ we include Lindblad operators defined in the basis that diagonalizes the instantaneous Hamiltonian $H_{\rm dr, \mbf k}(t)$
		\begin{align*}
			\Phi_{\mbf k} &= U_{\mbf k} \Psi_{\mbf k}\com
		\end{align*}
		where
		\begin{align}
			U_{\mbf k}&=\bbm 1 &0 &0 &0 \\ 0&1/\sqrt 2 &\tau_ze^{-i\tau_z\phi_k}/\sqrt 2 & 0 \\ 0 & \tau_z e^{i\tau_z\phi_k}/\sqrt 2 & -1/\sqrt 2 & 0 \\ 0 & 0&0&1\ebm\com\label{eq:unitaryTransform4x4}
		\end{align}
		and $\phi_{\mbf k}$ is defined via
		\begin{align*}
			\left|\mbf k+ e /\hbar \mbf A \right| e^{i\phi_{\mbf k}}=\vb{k_x+\frac{\rm e}{\hbar} A_x+i (k_y+\frac{\rm e}{\hbar}A_y)} \dt
		\end{align*}		
		In this basis we introduce 
		\begin{align*}
			L^\alpha&=\sqrt{c_\alpha} \bbm 
				0&\delta_{\alpha,1}&\delta_{\alpha,3}&0\\
				\delta_{\alpha,2}&0&\delta_{\alpha,5}&\delta_{\alpha,7}\\
				\delta_{\alpha,4}&\delta_{\alpha,6}&0&\delta_{\alpha,9}\\
				0&\delta_{\alpha,8}&\delta_{\alpha,10}&0
			\ebm \quad \text{for }\alpha=1,2,\dots,10\\
			L^{11}&=\sqrt{\gamma_z} \sigma_z
		\end{align*}
		with for now arbitrary constants $c_\alpha$. Here $c_5$ and $c_6$ correspond to decay effects and $\gamma_z$ corresponds to dephasing effects in the singly occupied sector. Additionally we explicitly allow for exchange of particles with the backgate. The time scale and dynamics for the exchange of particles are set by the damping constants $c_1-c_4$ and $c_7-c_{10}$.

		We note that the transformation in Eq.~\ref{eq:unitaryTransform4x4} is ill-defined when $\abs{\bk + e/\hbar \bA}=0$. In this case $H_\bk=0$ and the instantaneous Hamiltonian is diagonal with respect to any basis. We choose to implement the same Lindblad operators as above in the original $AB$-basis for this case.
		
		We find that the resulting equations of motion for the density matrix decouple into different sectors and write the density matrix in the sector that is relevant for computing the current as
		\begin{align*}
			\rho_{\mbf k}&= \sigma_g + \rho_{\mbf k,x} \sigma_x + \rho_{\mbf k,y} \sigma_y + \rho_{\mbf k,z} \sigma_z \eqb
			+ \rho_{\mbf k,0} \sigma_z^{(0)}+\rho_{\mbf k,2} \sigma_z^{(2)}
		\end{align*}
		The resulting equations of motion are
		\begin{widetext}
		\begin{align*}
			\hbar \del_t \rho_{\mbf k,x} &= \eb \rho_{\mbf k,z}-\ea \rho_{\mbf k,y} -\vsb{\Gamma +\vb{c_1+c_3+c_8+c_{10}}/2 } \rho_{\mbf k,x} \\
			\hbar \del_t \rho_{\mbf k,y} &= \ea \rho_{\mbf k,x}-\vsb{\Gamma +\vb{c_1+c_3+c_8+c_{10} }/2 } \rho_{\mbf k,y} \\
			\hbar \del_t \rho_{\mbf k,z} &= \eb \rho_{\mbf k,x} +c_3 \vb{1/2+\rho_{\mbf k,0}-\rho_{\mbf k,z}  } -c_4 \rho_{\mbf k,2} 
			+ c_5 \vb{1/2+ \rho_{\mbf k,0}-\rho_{\mbf k,z}} -c_6 \vb{1/2+ \rho_{\mbf k,z}-\rho_{\mbf k,2} }\eqb  
			-c_7 \rho_{\mbf k,0} -c_8 \vb{1/2+\rho_{\mbf k,z} - \rho_{\mbf k,2} }  \\
			\hbar \del_t \rho_{\mbf k,0} &= -(c_7+c_9)  \rho_{\mbf k,0} - c_{10} \vb{1/2 +\rho_{\mbf k,0}-\rho_{\mbf k,z} }- c_8 \vb{1/2+ \rho_{\mbf k,z}-\rho_{\mbf k,2} }  \\
			\hbar \del_t \rho_{\mbf k,2}&= - (c_2+c_4)  \rho_{\mbf k,2} + c_3 \vb{1/2+\rho_{\mbf k,0}-\rho_{\mbf k,z}} + c_1 \vb{1/2+\rho_{\mbf k,z}-\rho_{\mbf k,2}  } \com
		\end{align*}
		\end{widetext}
		where
		\begin{align*}
			\gamma_2&=(c_5+c_6)/2+2\gamma_z\\
			\ea&=2\tau_z v_{\rm F} \vsb{ \hbar \abs {\mbf k} + e \mbf k \cdot \mbf A/\abs{\mbf k} } \nonumber\\
			\eb&=2\tau_z v_{\rm F} \vsb{ e \mbf A \times \mbf k / \abs{\mbf k} } \nonumber \dt
		\end{align*}
		We note that while we give the equations of motion in the basis diagonalizing $H_{0,\bk}$ here, we implement them in the original $AB$-basis in the numerical simulations.

		We choose the damping constants Boltzmann distributed
		\begin{align*}
			\gamma_2&=1/T_2\\
			c_5 &= c_6 \exp(-2\beta \epsilon) & c_5+c_6&=1/T_1\\
			c_1 &= c_2 \exp(-\beta(-\epsilon-\mu)) & c_1+c_2&=1/T_p\\
			c_3 &= c_4 \exp(-\beta(\epsilon-\mu))& c_3+c_4&=1/T_p\\
			c_7 &= c_8 \exp(-\beta(\epsilon-\mu))& c_7+c_8&=1/T_p\\
			\text{and } c_9 &= c_{10} \exp(-\beta(-\epsilon-\mu)) & c_9+c_{10}&=1/T_p \com
		\end{align*}
		where $\epsilon=v_{\rm F} \sqrt{\vb{\hbar k_x+e A_x}^2+\vb{\hbar k_y+e A_y}^2} $ are the instantaneous eigenenergies. 

		This ensures that the ground state of the system without the light field is Fermi distributed with chemical potential $\mu$ and inverse temperature $\beta=1/(k_BT)$. 

		Note that $T_1$ and $T_2$ are the commonly-introduced decoherence measures. In analogy we define a third time scale $T_p$ for the exchange of particles with the backgate.	

		We solve the master equation numerically and then compute the current for each momentum point
		\begin{align*}
			\mbf j_{\mbf k} &= \ave{\ddel{H_{\mbf k}}{\mbf A}} =  ev_{\rm F} \vb{\tau_z \ave{\sigma_x} \mbf e_x + \ave{\sigma_y} \mbf e_y} \com
		\end{align*}
		where Pauli matrices here refer to the singly-occupied sector and empty and doubly occupied modes do not contribute to the current.	The conductivity is then obtained as $\sigma_{\rm xy}(\mbf k)=\lim_{E_{L} \rightarrow 0} \mbf j_{y, \mbf k}/E_{\rm L}$. We perform the calculation of $\mbf j_{y, \mbf k}$ at experimentally realistic values of $E_{L}$, and have checked that these values realize the linear response limit. Finally we define the conductivity density
		\begin{align*}		
			\tilde \sigma_{\rm xy}&=\frac 1A\, \sigma_{xy}(\mbf k) 
		\end{align*}
		where $A$ is the lattice size and the full conductivity
		\begin{align*}
			\sigma_{\rm xy}&= \sum_{\mbf k} \tilde \sigma_{\rm xy}\dt
		\end{align*}

		We note that a similar method for the calculation of current has been used in Ref.~\cite{sato_microscopic_2019}. The crucial difference is that we explicitly allow for the exchange of particles by including the empty and the fully occupied mode. In particular this also implies that we introduce a separate time scale for particle-exchange processes $T_p$. Also in our case the trace of the density matrix is ensured to be $1$ at all times as in Ref.~\cite{sato_microscopic_2019}. Quantitatively the approach presented here yields better agreement to experimental data from Ref.~\cite{mciver_light-induced_2019}. Furthermore including the empty and doubly-occupied mode is crucial for the calculation of the single-particle correlation function.

	\section{Rotating wave approximation for graphene: Rabi-Bloch bands}\label{app:rabi}
		We start from the graphene Hamiltonian from Eq.~\ref{eq:grapheneHamiltonian} with no longitudinal field
		\begin{align*}
			H_{\mbf k}&=H_{0,\mbf k}+H_{\rm dr,\mbf k}(t) \dt
		\end{align*}
		The undriven Hamiltonian is diagonalized by
		\begin{align*}
			H_0^d&=U^\dag H_0 U = \hbar v_{\rm F} k \sigma_z\\
			U &= 1/\sqrt{2} \bbm 1&1 \\e^{i\phi_k} & -e^{i\phi_k} \ebm \\
			e^{i\phi_k} &= \frac{\tau_z k_x + i k_y}k \dt
		\end{align*}
		In this basis the driving Hamiltonian is
		\begin{align*}
			H_{\rm dr,\mbf k}(t)&= \frac{eE_{\rm dr} v_{\rm F}}{\omega_{\rm dr}} 
				\bbm 
					\tau_z s_{\rm dr} &
					-i \sigma_{\rm pol} c_{\rm dr}\\
					i \sigma_{\rm pol} c_{\rm dr}
					&-\tau_z s_{\rm dr}\\
				\ebm
		\end{align*}
		where 
		\begin{align*}
			s_{\rm dr}&=\sin(\omega_{\rm dr}t-\tau_z \sigma_{\rm pol} \phi_{\rm k})\\
			c_{\rm dr}&=\cos( \omega_{\rm dr} t- \tau_z \sigma_{\rm pol} \phi_{\rm k})\dt
		\end{align*}
		Next we do the rotating wave approximation, keeping only those terms, non-oscillatory in the rotating frame. Then
		\begin{align*}
			H_{\rm dr,\mbf k}(t) &\approx \frac{ \sigma_{\rm pol} e E_{dr}v_{\rm F}}{2 \omega_{\rm dr}} 
			\bbm 
				0 \quad -ie^{-i \omega_{\rm dr}t+i \tau_z \sigma_{\rm pol} \phi_k}  \\  
				ie^{i \omega_{\rm dr}t-i \tau_z \sigma_{\rm pol} \phi_k} \quad 0 \ebm \dt
		\end{align*}
		In analogy to the Rabi problem the system can now be solved analytically. The eigenenergies are $E_{\rm R,\pm}=-\hbar \omega_{\rm dr}/2\pm \hbar \Omega_{\rm R}  $ and the eigenstates is
		\begin{align*}
			|\psi(t)\rangle &= |\psi_{\rm R,+}(t)\rangle e^{-i\Omega_{\rm R}t} + |\psi_{\rm R,-}(t)\rangle e^{i\Omega_{\rm R}t} \com
		\end{align*}
		where
		\begin{align*}
			|\psi_{R,\pm}(t)\rangle&=	\bbm -i \sigma_{\rm pol} e^{i \tau_z \sigma_{\rm pol} \phi_{\rm k}} \vb{a\mp \frac{a \Delta-b\lambda}{\Omega_R}} e^{-i \omega_{\rm dr} t/2} \\ \vb{b \pm \frac{b\Delta +a\lambda }{\Omega_R}} e^{i \omega_{\rm dr} t/2} \ebm \dt
		\end{align*}
		Furthermore
		\begin{align*}
			\Delta&=\frac{\omega_{\rm dr}-2v_{\rm F} k}{2}\\
			\Omega_{\rm R}&=\sqrt{\lambda^2+\Delta^2}\\
			\lambda&= \frac{e E_{\rm dr} v_{\rm F}}{2 \hbar \omega_{\rm dr}}
		\end{align*}
		and the constants $a$ and $b$ are integration constants constrained by normalizing $|\psi_{\rm R,\pm}(t)\rangle$. The remaining freedom in $a$ and $b$ determines the initial state.
		The band velocity in the $y$-direction is immediately obtained as
		\begin{align*}
			v_y^\pm(\mbf k)&= \del_y E_{\rm R,\pm}
		\end{align*}
		Further note that 
		\begin{align*}
			\sigma_{\rm pol} e^{i \tau_z \sigma_{\rm pol} \phi_k}&= \frac{\tau_z (\sigma_{\rm pol} k_x+i k_y)}{k}\dt
		\end{align*}	
		Next we determine the instantaneous Berry curvature. For this we need the eigenstates with respect to the original $AB$-basis
		\begin{align*}
			|\psi^{AB}_{\rm R,\pm}(t)\rangle&= U |\psi_{\rm R,\pm}(t)\rangle
		\end{align*}
		The Berry connection is now given by
		\begin{align*}			
			A^{\pm}_j&=i \ave{\psi^{AB}_{\rm R,\pm}|\del_j|\psi^{AB}_{\rm R,\pm}}
		\end{align*}
		and as a result we obtain the Berry curvature as
		\begin{align*}
			&\Omega^\pm(\mbf k)= \del_y A_x - \del_x A_y \\
			&=\pm \frac{\sigma_{\rm pol}v_{\rm F} \lambda^2}{2k\Omega_{\rm R}^3} \pm \frac{{\rm Re}\vsb{(k_y+i \sigma_{\rm pol} k_x) e^{i \omega_{\rm dr} t}}v_{\rm F} \lambda \Delta}{2 k^2 \Omega_{\rm R}^3} \label{eq:RabiBerryCurv} \dt
		\end{align*}
		Finally we can compute the resulting Hall conductivity from Eq.~\ref{eq:relationCurvatureCond} from the main text, where the Rabi occupations are computed from the density matrix $\rho_\mbf k$ as
		\begin{align*}
			n_{\rm R,\pm}(\mbf k)&= \ave{\psi^{AB}_{\rm R,\pm}|\rho_{\mbf k}|\psi^{AB}_{\rm R,\pm}}\dt
		\end{align*}
		
		Note that both the Berry curvature and the occupations are time dependent. Therefore there is a time-independent contribution from time-dependent curvature and occupations. We have checked that this contribution is at least an order of magnitude smaller than the time-independent contribution and hence the quantities can be averaged independently $\overline{\Omega^\pm(\bk,t)n_{\rm R,\pm}(\mbf k,t)}\approx\overline{\Omega^\pm(\bk,t)}\;\overline{n_{\rm R,\pm}(\mbf k,t)}$. Hence we can drop the second, time-dependent contribution to the Berry curvature.

	\section{Numerical results within the Rabi approximation} \label{app:shiftingRabi}
		\begin{figure}[tb]
			\centering
			\includegraphics[width=\linewidth]{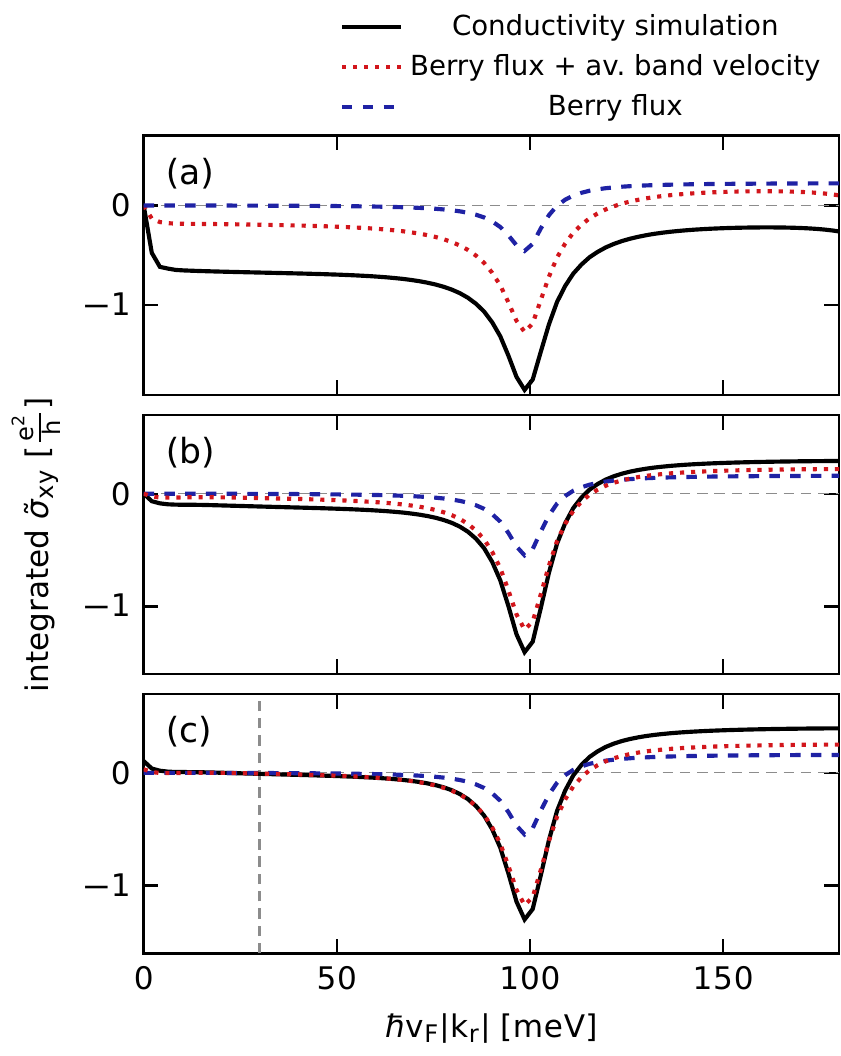}
			\caption{We depict the conductivity density, integrated over a disk in momentum space of radius $\bk_{\rm r}$. For comparison, we show the momentum resolved Berry flux, integrated over the same disk in momentum space, and the sum of the integrated Berry flux and the integrated average band velocity. Both are obtained within the Rabi approximation.
			In panel (c) the curves have been shifted such that their value vanishes at $\hbar \rm v_F \abs{k_{\rm r}}=\usi{30}{\milli\eV}$ for a better comparison of the contribution of the first resonance. 				
			In all plots we use $E_{\rm dr}=\usi{3}{\mega\volt\per\meter}$, $\omega_{\rm dr}=\usi{2\pi\cdot 48}{\tera\hertz}\approx\usi{200}{\milli\eV}/\hbar$, $T=\usi{80}{\kelvin}$, $E_{\rm L}=\usi{0.84}{\kilo\volt\per\meter}$, $\mu=0$ and the driving pulse is ramped with a tanh over $\usi{1}{\pico\second}$. Panel (a) shows $T_1=\usi{50}{\pico\second}$, $T_2=\usi{10}{\pico\second}$, $T_p=\usi{20}{\pico\second}$, while panels (b) and (c) show $T_1=\usi{4}{\pico\second}$, $T_2=\usi{0.8}{\pico\second}$, $T_p=\usi{1.6}{\pico\second}$.}
			\label{fig:app:unshiftedRabi}
		\end{figure}
		As described in the main text, mapping graphene onto the Rabi problem is a good approximations close to the first resonance. By the nature of the approximations made the Rabi results are not valid close to the Dirac point and hence the contribution of the Dirac point can not be captured. We have therefore shifted the curves in Fig.~\ref{fig:rabiBerryCurv}(b), such that only the conductivity density of the first resonance is integrated. For completeness we show the unshifted version in Fig.~\ref{fig:app:unshiftedRabi}(a). For the low dissipation considered the Dirac point obtains a significant contribution that is larger than the contribution of the first resonance. We also show results for higher dissipation in Fig.~\ref{fig:app:unshiftedRabi}(b) and (c), where the contribution of the Dirac point is small.

	\section{Calculation of single-particle correlation function}\label{app:singleParticleCorrelation}		
		Given a density matrix at a certain time $\rho(t_1)$ by the numerical methods of Appendix A, the single-particle correlation function $\braket{c(t_2)^\dagger c(t_1)}$ can be calculated. The state is acted upon with an annihilation operator $c$ which gives a matrix of the shape
		\begin{equation*}
		c\rho(t_1) = \begin{pmatrix}
		0 & 0 & 0 & 0 \\
		r_1 & 0 & 0 & 0 \\
		r_2 & 0 & 0 & 0 \\
		0 & r_3 & r_4 & 0 \\
		\end{pmatrix}.
		\end{equation*}
		This object is evolved to a later time $t_2$ using the equations of motion
		\begin{align*}
		\dot{r}_1 &= 
		        -r_1 (i \mu +\Gamma_1+\Gamma_3+\Gamma_4)
		        -e^{-i\phi}r_2(+i \nu_F |q| + \Gamma_3-\Gamma_4)\\
		\dot{r}_2 &=
		        -r_2 (i \mu +\Gamma_1+\Gamma_3+\Gamma_4)
		        -e^{+i\phi}r_1(+i \nu_F |q| + \Gamma_3-\Gamma_4)\\
		\dot{r}_3 &=  
		        -r_3(i \mu +\Gamma_2+\Gamma_3+\Gamma_4)
		        -e^{+i\phi}r_4(-i \nu_F |q| +\Gamma_3-\Gamma_4)\\
		\dot{r}_4 &= 
		        -r_4(i \mu +\Gamma_2+\Gamma_3+\Gamma_4)
		        -e^{-i\phi}r_3(-i \nu_F |q| +\Gamma_3-\Gamma_4),
		\end{align*}
		where 
		\begin{align*}
		\Gamma_1 &= \frac{1}{2}(\gamma_-^{(ud)}+\gamma_-^{(ld)}+\gamma_z)&
		\Gamma_2 &= \frac{1}{2}(\gamma_+^{(uu)}+\gamma_+^{(lu)}+\gamma_z)\\
		\Gamma_3 &= \frac{1}{4}(\gamma_-^{(b)}+\gamma_-^{(uu)}+\gamma_+^{(ud)})&
		\Gamma_4 &= \frac{1}{4}(\gamma_+^{(b)}+\gamma_-^{(lu)}+\gamma_+^{(ld)}).
		\end{align*}
		Here it is $q=q_x+i q_y$ the momentum and $\phi=\mathrm{arg}(q)$ its phase. $\mu$ is the chemical potential.

		At any time $t_2$ this state can be acted upon with $c^\dagger$ and traced over to give the correlation function. 

		The occupations of the system are then calculated using an approach inspired by trARPES \cite{freericks_theoretical_2009}
		\begin{align*}
		n(\bk,\omega) &= \frac{1}{t-t_0}\int_{t_0}^t\int_{t_0}^t \braket{c^\dagger(t_2) c(t_1)}e^{i\omega(t_2-t_1)}dt_1 dt_2 \\
		&=\frac{2}{t-t_0}\mathrm{Re}\left[
		\int_{t_0}^t\int_{t_1}^t \braket{c^\dagger(t_2)c(t_1)}e^{i\omega(t_2-t_1)}dt_2dt_1
		\right].
		\end{align*}

		The occupations of the individual Floquet bands are assigned by integrating $n(\bk,\omega)$ over frequency intervals of multiples of $\omega_d/2$, starting and ending centered at the band gaps.
		\begin{equation*}
		n_{\alpha,\pm} (\bk) = \int_{(\alpha-\frac{1}{4}\pm\frac{1}{4}) \omega_d}^{(\alpha+\frac{1}{4}\pm\frac{1}{4})\omega_d} n(\bk,\omega) d\omega
		\end{equation*}
		The effective Floquet band occupations are found by summation over the Floquet index:
		\begin{equation*}
		n_\pm(\bk)=\sum_\alpha n_{\alpha,\pm}(\bk)\dt
		\end{equation*}	

	\section{Floquet-Berry-curvature calculation}\label{app:floquetBerry}		
		Here we present the details on the calculation of the Berry curvature of Floquet bands. For each momentum $\mbf k$ we use the quasi-energy operator in the extended Floquet-Hilbert space (for details see for example \cite{eckardt_high-frequency_2015})
		\begin{align*}
			Q &=\bbm \ddots & \vdots &\vdots & \reflectbox{$\ddots$} \\
				\cdots & H_0  & H_1 & \cdots \\
				\cdots & H_{-1} & H_0+\hbar \omega_{\rm dr} & \cdots \\
				\reflectbox{$\ddots$}&\vdots&\vdots&\ddots
			\ebm
		\end{align*}
		where 
		\begin{align*}
			H_m&=\int \D t\; e^{-im \omega_{\rm dr}t} (H_{0,\bk} + H_{{\rm dr,\bk}}(t)) \text{ and}\\
			g_{\rm env}(t)&=1 \dt
		\end{align*}
		In order to get the Floquet eigenstates and eigenenergies we diagonalize $Q$ after truncating such that $-4\leq m \leq 4$. The Floquet band structure is obtained from combining the eigenenergies of different momentum points. Subsequently we use the method presented in \cite{takahiro_fukui_chern_2005} in order to determine the Berry curvature numerically. Finally the Floquet band velocity is obtained by numerically computing the momentum derivative of the Floquet eigenenergies.

	\section{Weyl gauge}\label{app:gaugeChoice}
		Here we discuss the meaning of different gauge choices and their importance for our method. For all our calculations we choose the Weyl gauge, i.e. we choose the scalar potential $\phi=0$ and the time-dependent vector potential $\mbf A(\mbf r,t)=-\int \D t\; \mbf E(\mbf r,t)$. Using the Peierls Substitution $\mbf k \To \mbf k - \frac q \hbar \mbf A(\mbf r,t)$ this leads to time-dependent shift of the momentum in the Hamiltonian. This can be viewed as a time-dependent shift of the band structure. The Weyl gauge is particularly useful for an electric field that is spatially constant within the graphene sheet in the $x$-$y$-plane. In this case the vector potential within the $x$-$y$-plane can also be chosen independent of position and hence the contribution to the Hamiltonian decouples in momentum space. 
		As an example for the choice of gauge we consider a uniform electric field $\mbf E= E \hat{\mbf e}_x$ and vanishing magnetic field $\mbf B=0$. For this case the Weyl gauge implies $\mbf A=-\mbf E t$. The vector potential is indeed independent of position. An alternative gauge choice would be a special case of the Coulomb gauge, $\mbf A=0$. This choice implies $\phi=E x$ which can be viewed as a tilt of the lattice potential. The resulting Hamiltonian obtains a non-trivial spatial dependence and hence is not diagonal any more in momentum space.

	\section{Hall conductivity for a cut along the y-direction} \label{app:conductivityYCut}
		\begin{figure}[tb]
			\centering
			\includegraphics[width=\linewidth]{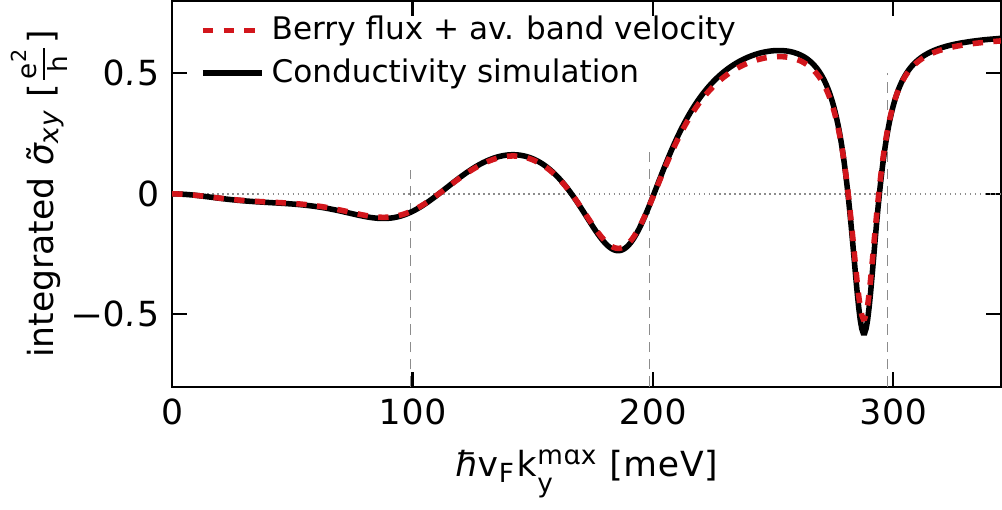}
			\caption{We show the conductivity density $\tilde \sigma_{\rm xy}$ integrated over momenta in an interval on the $y$-axis, specifically fulfilling $k_x=0, \abs{k_y}<k_y^{\rm max}$. 
			This integrated conductivity agrees with the sum over the Berry flux and the average band velocity, integrated over the same interval.
			The parameters are $E_{\rm dr}=\usi{26}{\mega\volt\per\meter}$, $\omega_{\rm dr}=\usi{2\pi\cdot 48}{\tera\hertz}\approx\usi{200}{\milli\eV}/\hbar$, $T_1=\usi{1}{\pico\second}$, $T_2=\usi{0.2}{\pico\second}$ and $T_p=\usi{0.4}{\pico\second}$, $T=\usi{80}{\kelvin}$ and $E_{\rm L}=\usi{1.7}{\kilo\volt\per\meter}$. All observables are shown after a steady state is achieved for a tanh-type ramp of the driving field strength.}
			\label{fig:app:yCutConductivity} 
		\end{figure}
			In Fig.~\ref{fig:app:yCutConductivity} we display the momentum resolved contributions to the conductivity, integrated over the momentum state interval from $-k_{y}^{\rm max}$ to $k_{y}^{\rm max}$ on the $k_y$- axis. The corresponding integral over the contributions to $n\bar v/E_{\rm L} + \Phi_{\rm xy}$ are depicted as well. We find essentially perfect agreement for these quantities. Hence, the deviations between these quantities in Fig.~\ref{fig:compFloquetFluenceDep}(a) in the main text arise predominantly from the $k_x$-direction, where the averaged band velocity vanishes and only the Berry flux contributes.

	\section{Subsequent opening of gaps}\label{app:gapOpening}

		For larger electric field strength it is no longer sufficient to consider the first resonance only. The contribution in the high-frequency limit without damping has been analyzed in Ref.~\cite{oka_photovoltaic_2009,kitagawa_topological_2010,kitagawa_transport_2011,dehghani_out--equilibrium_2015,mikami_brillouin-wigner_2016}. In this limit there are no resonant contributions and the total Hall current is $\sigma_{xy}=-2 \frac{e^2}{h}$. For this result it is assumed that only the lower graphene band is occupied. Under experimental conditions finite frequency driving leads to excitations into the upper graphene band. For intermediate driving strength the Berry curvature is still well localized around the Dirac point and individual resonances. We can therefore investigate each of the contributions separately. Depending on the strength of damping and dephasing effects one obtains a steady state with significant occupation in the upper graphene band close to the Dirac point, see Fig.~\ref{fig:floquetBands}(a). The upper band has opposite Berry curvature and hence contributes to the Hall current with opposite sign. Hence the Hall current arising from the Dirac point is significantly reduced for experimental conditions. Since the occupation of the lower band is always larger than the one of the upper band the net contribution from the Dirac point is always negative. 

\begin{figure}[htb]
			\centering
			\includegraphics[width=\linewidth]{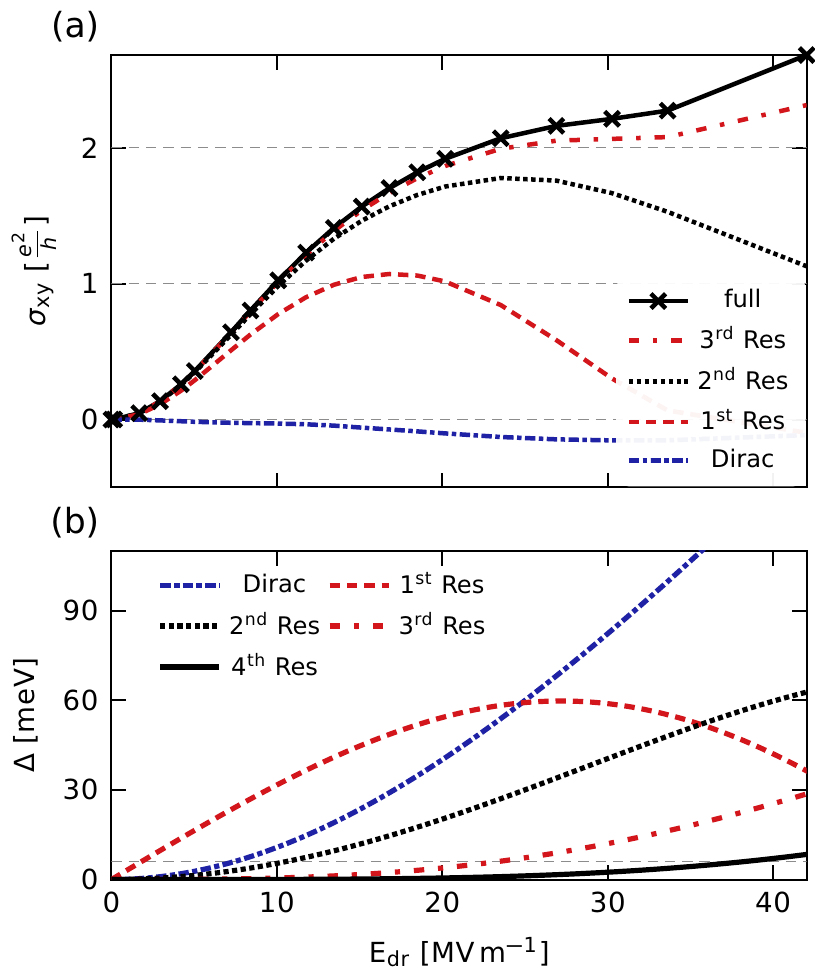}
			\caption{(a) Resonance resolved conductivity dichroism as a function of electric-field strength. The solid black line shows the full current, while the dash-dotted blue line shows only the contribution from the gap at the Dirac point and other lines show the contributions up to and including the n-th resonance as indicated in the legend. (b) Gap sizes as a function of electric-field strength. The dash-dotted blue line shows the gap at the Dirac point, while other lines show the gaps at the n-th resonance as indicated in the legend. The dashed gray line shows the approximate scale of temperature, damping and dephasing effects $k_BT\approx\hbar/T_1\approx\usi{6}{\milli\eV}$. The parameters used are $\omega_{\rm dr}=\usi{2\pi\cdot 48}{\tera\hertz}\approx\usi{200}{\milli\eV}/\hbar$, $T_1=\usi{100}{\femto\second}$, $T_2=\usi{20}{\femto\second}$, $T_p=\usi{40}{\femto\second}$ $T=\usi{80}{\kelvin}$, $E_{\rm L}=\usi{1.7}{\kilo\volt\per\meter}$, $\mu=0$ and the envelope of the driving pulse is a tanh-type interpolation from $0$ to $1$, that reaches $1$ after $\usi{1}{\pico\second}$.}
			\label{fig:gapResolvedCurrent}
		\end{figure}

		The main resonant contribution comes from those gaps that are lying on the original lower Dirac cone. For these gaps the Floquet band below the gap has positive curvature while the band above has equal and opposite curvature. Hence for equal occupation of both bands close to the gap, there is no net contribution to the current. This is the case for higher order gaps with magnitude smaller than temperature and damping. We say that these gaps are closed, see Fig.~\ref{fig:gapResolvedCurrent}(b). In Fig.~\ref{fig:gapResolvedCurrent} only the first gap is open for electric field strengths smaller than $\usi{8}{\mega\volt\per\meter}$. In this regime the current is well described by the Rabi-Berry curvature. For field strengths larger than $\usi{10}{\mega\volt\per\meter}$ we expect the current arising from the first resonance to saturate. The reduction that can be seen in this regime in Fig.~\ref{fig:gapResolvedCurrent} is a numerical artifact that we explain in App.~\ref{app:resonanceBroadening}. While the current arising from the first resonance saturates, the second resonance gap opens and for higher $E_{\rm dr}$ leads to a further increase of the Hall conductivity. At even higher field strengths higher-order gaps open subsequently. For each gap the net contribution to the current is positive since there is more occupation in the band below the gap than in the one above. Hence the total resonant contribution is opposite to the high-frequency contribution. Furthermore we find numerically that the magnitude of the high-frequency contribution is always smaller than the magnitude of the resonant contributions and usually is a minor effect. This is in agreement with the sign of the current in Ref.~\cite{mciver_light-induced_2019}.

	\section{Resonance-resolved conductivity and resonance broadening}\label{app:resonanceBroadening}
\begin{figure}[tb]
			\centering
			\includegraphics[width=\linewidth]{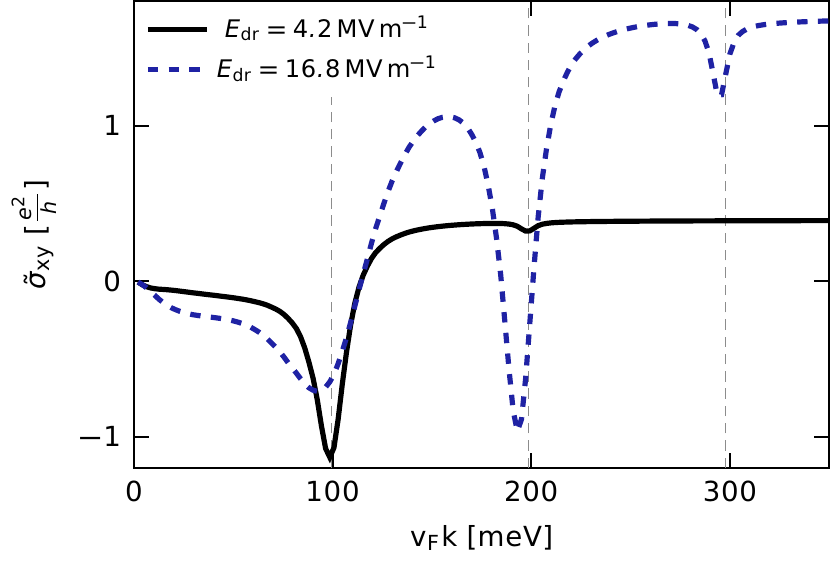}
			\caption{Integrated Hall conductivities. We show the sum of the conductivity for all momenta smaller than a threshold value as a function of threshold momentum. The first, second and third resonance are indicated by dashed lines. We use $\omega_{\rm dr}=\usi{2\pi\cdot 48}{\tera\hertz}\approx\usi{200}{\milli\eV}/\hbar$, $T_1=\usi{1}{\pico\second}$, $T_2=\usi{200}{\femto\second}$, $T_p=\usi{200}{\femto\second}$ $T=\usi{80}{\kelvin}$, $E_{\rm L}=\usi{1.7}{\kilo\volt\per\meter}$, $\mu=0$ and the envelope of the driving pulse is a tanh-type interpolation from $0$ to $1$, that reaches $1$ after $\usi{1}{\pico\second}$.}
			\label{fig:momentumResolvedDifEd}
		\end{figure}

		For low driving field strength $E_{\rm dr}$ the current is well localized around individual resonances. In contrast, for large $E_{\rm dr}$ resonances start overlapping and it is therefore difficult to identify the current arising from individual resonances. We show an example of this phenomenon in Fig.~\ref{fig:momentumResolvedDifEd}. For low values of $E_{\rm dr}$ there is no contribution to the current in between resonances. Hence the integrated conductivity shown in Fig.~\ref{fig:momentumResolvedDifEd} is constant. For larger values of $E_{\rm dr}$ resonances get broadened and there is no such constant regime. This is an indication that the contribution of neighboring resonances is now overlapping. Since the contribution from resonances is always negative below and positive above the resonance, overlapping resonances lead to canceling contributions. 

		When we compute the resonance-resolved conductivity as in Fig.~\ref{fig:gapResolvedCurrent}, we do this by integrating the current up to the momentum value half-way in between the corresponding resonances. In other words we use the corresponding value of the curve in Fig.~\ref{fig:momentumResolvedDifEd}. Once resonances start overlapping the contributions cancel and hence lead to decreasing contributions of the inner resonances. This is the reason why the curves in Fig.~\ref{fig:gapResolvedCurrent} decrease.

	\section{Chemical potential dependence at low fluence}
		The momentum-resolved conductivity allowed us to identify the different contributions to the transverse current. In experiment, however, such data is not easily accessible. Instead it is possible to tune the applied backgate, i.e. the chemical potential \cite{mciver_light-induced_2019}. When increasing the chemical potential, momenta close to the Dirac point are fully occupied and, due to Pauli blocking, do not contribute to the conductivity. For momentum modes smaller than the first bare resonance negative contributions to the conductivity dominate. For increasing chemical potential conductivity from these modes becomes suppressed and the total conductivity increases. Near the first resonance the situation reverses. Now momentum modes above the resonance become fully occupied and increasing the chemical potential further leads to decreasing total conductivity. Hence the chemical-potential-resolved transverse conductivity shows a clear signature of the resonant behavior. \vspace*{0.6cm}		

	\section{Comparison of fixed chemical potential and fixed density.}
		For the simulation of the experiment from Ref.~\cite{mciver_light-induced_2019} it is crucial to work at fixed chemical potential instead of fixed density. To illustrate the difference we show a simulation enforcing fixed density for each momentum mode during the time evolution in Fig.~\ref{fig:app:fluenceComp}(a). The parameters are the same as in Fig.~2(a) in the main text. The shape of the curve is fundamentally different from the experimental data. We conclude that the exchange of electrons with non-illuminated regions of the graphene sample as well as with the substrate is important even on the short time scales of the circularly polarized pulse. 

	\section{Fluence dependence}\label{app:fluenceDependence}
		
		For completeness and for better comparison to the experiments in Ref.~\cite{mciver_light-induced_2019} we show the data from Fig.~2(b) in the main text as a function of fluence instead of peak driving field in Fig.~\ref{fig:app:fluenceComp}(b).

		\begin{figure*}[p]
			\centering
			\includegraphics[width=\linewidth]{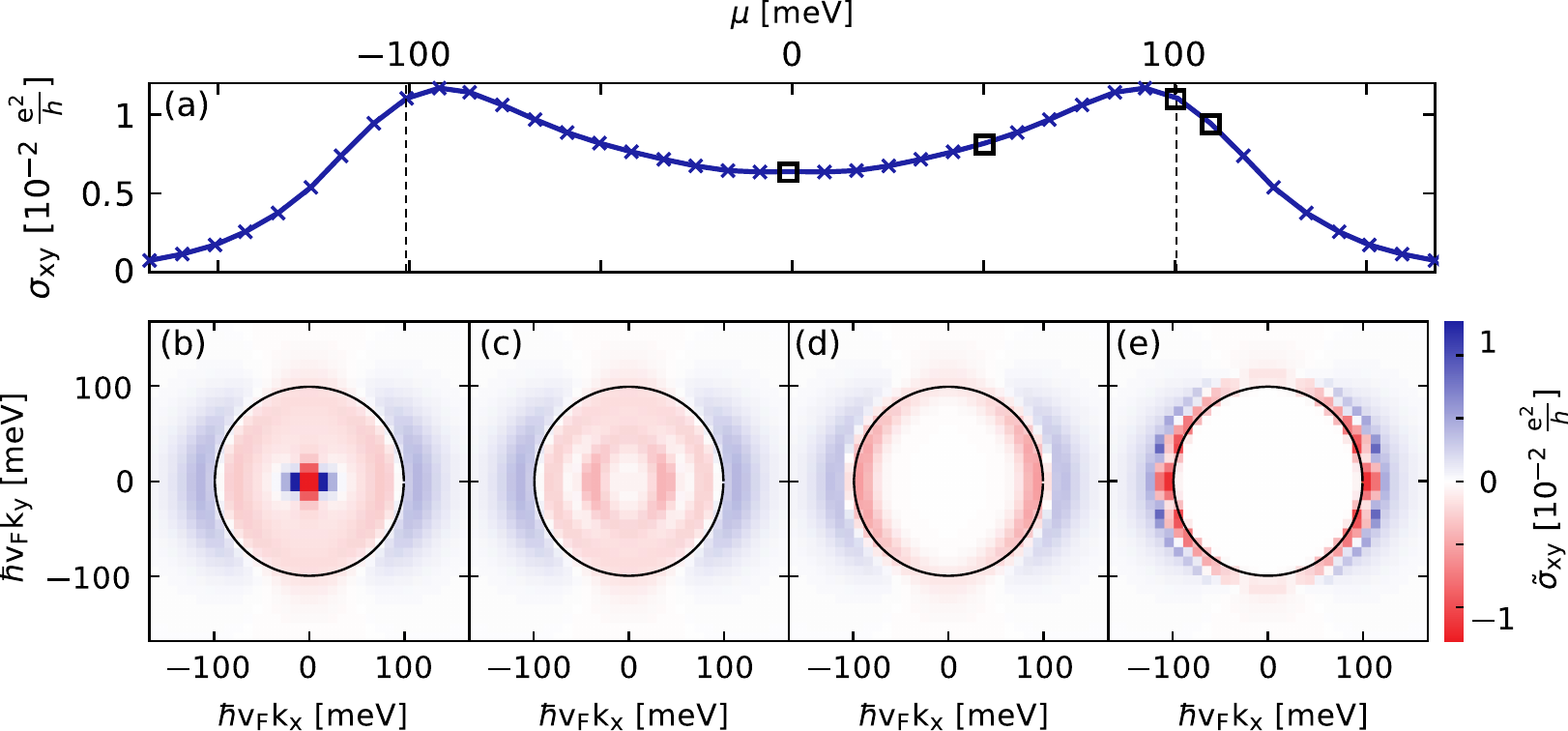}
			\caption{Circular dichroism of the transverse conductivity. Panel (a) shows the total conductivity as a function of applied chemical potential. Panels (b)-(e) show the momentum resolved conductivity for several different chemical potentials as indicated by black squares in panel (a). The panels are aligned in the same order as the squares in panel (a). Black circles denote the position of the first bare resonance $2v_{\rm F}k=\omega_{\rm dr}$. For these panels we average the conductivity of opposite momentum modes $\overline{\sigma}_y(\mbf k)=\vsb{\sigma_y(\mbf k)+\sigma_y(-\mbf k)}/2$ and for both Dirac points. The center of each panel $(0,0)$ is positioned at the Dirac point. The parameters for all panels are $E_{\rm dr}=\usi{1}{\mega\volt\per\meter}$, $\omega_{\rm dr}=2\pi\cdot \usi{48}{\tera\hertz}$, $T_1=\usi{100}{\femto\second}$, $T_2=\usi{20}{\femto\second}$, $T_p=\usi{25}{\femto\second}$, $T=\usi{80}{\kelvin}$, $E_{\rm L}=\usi{840}{\volt\per\meter}$ and $g_{\rm env}(t)$ is a Gaussian envelope with full width have maximum $t_{\rm FWHM}=\usi{1}{\pico\second}$.}
			\label{fig:lowIntensity}
		\end{figure*}

		\begin{figure*}[p]
			\centering
			\includegraphics[width=\linewidth]{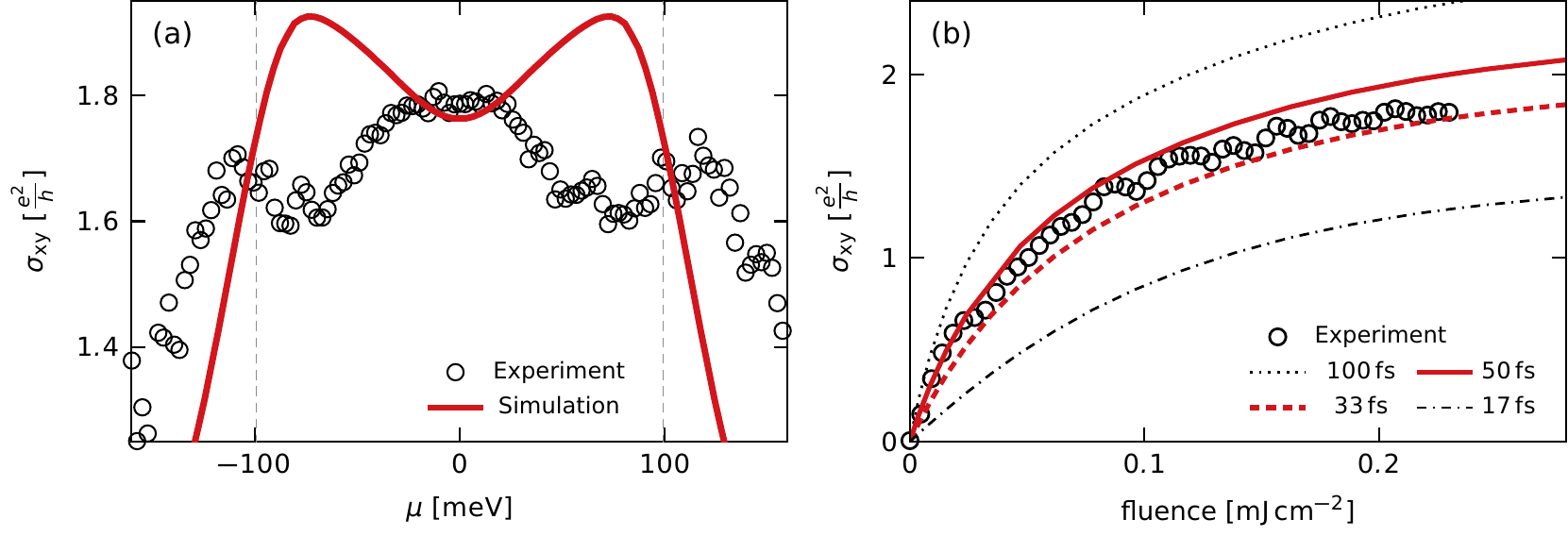}
			\caption{(a) Chemical potential resolved circular dichroism of the transverse conductivity. Black circles show experimental data from \cite{mciver_light-induced_2019} and the solid red line shows results from our numerical simulation. Here $\mu$ is the chemical potential of the initial state and we do not allow for the exchange of particles during the simulation. 
			(b) Fluence dependence of the current dichroism for several values of the particle-exchange time-scale $T_p$, as indicated in the legend. We see that a value in the range of $T_p=\usi{30-50}{\femto\second}$ is consistent with the experiment. For both panels the parameters for the numerical simulation are $\omega_{\rm dr}=\usi{2\pi\cdot 48}{\tera\hertz}\approx\usi{200}{\milli\eV}/\hbar$, $T_1=\usi{100}{\femto\second}$, $T_2=\usi{20}{\femto\second}$, $T=\usi{80}{\kelvin}$, $E_{\rm L}=\usi{1.7}{\kilo\volt\per\meter}$ and the driving pulse has a Gaussian envelope with electric-field strength FWHM of $\usi{\sqrt 2}{\pico\second}$, corresponding to intensity FWHM of $\usi{1}{\pico\second}$. Furthermore we use $E_{\rm dr}=\usi{26}{\mega\volt\per\meter}$ and $T_p=\usi{36}{\femto\second}$ in panel (a) and $\mu=0$ in panel (b).
			For details on the experimental data see \cite{mciver_light-induced_2019}.}
			\label{fig:app:fluenceComp}	
		\end{figure*}

	\end{appendix}

\bibliography{bib_paper}

\end{document}